\documentclass[12pt]{iopart}

\usepackage{iopams} 
\usepackage{dsfont}
\usepackage{tikz}

\newcounter{mycounter}
\usepackage{calc}
\DeclareFontFamily{OT1}{pzc}{}
\DeclareFontShape{OT1}{pzc}{m}{it}{<-> s * [1.100] pzcmi7t}{}
\DeclareMathAlphabet{\mathpzc}{OT1}{pzc}{m}{it}
\DeclareMathAlphabet{\mathpzc}{OT1}{pzc}{m}{it}
\usepackage[super,compress]{cite}
\usepackage{graphicx}
\usepackage{xcolor}
\expandafter\let\csname equation*\endcsname\relax
\expandafter\let\csname endequation*\endcsname\relax
\usepackage{amsmath}
\usepackage[utf8]{inputenc}
\usepackage{tikz-cd}
\usepackage{listings}
\usepackage[perpage]{footmisc}

\begin{document}
\title[Alternative for Black Hole Paradoxes]{Alternative for Black Hole Paradoxes}
\author{Reinoud Jan Slagter}
\address{ASFYON, Astronomisch Fysisch Onderzoek Nederland, The Netherlands\\
former: University of Amsterdam, The Netherlands}
\ead{info@asfyon.com}
\vspace{10pt}
\begin{indented}
\item[]June 2022
\end{indented}
\begin{abstract}
A throughout investigation is made of the exact black hole solution in five-dimensional warped conformal dilaton gravity, found in an earlier investigation. The singularities of the dynamical black hole spacetime are determined by the zeros of a meromorphic quintic polynomial, which has no essential singularities.
The solutions of the polynomial are analyzed in the complex plane in relation to the icosahedron group and by the Hopf-fibrations of the Klein surface.
The model fits the antipodal boundary condition, i.e., antipodal points in the projected space are identified using the embedding of a Klein surface in $\mathds{C}^2$, using the $\mathds{Z}_2$ symmetry on the two sides of the brane.
If one writes $^{(5)}g_{\mu\nu}=\omega^{4/3}{^{(5)}}{\tilde g_{\mu\nu}}, {^{(5)}}{\tilde g_{\mu\nu}}={^{(4)}}{\tilde g_{\mu\nu}}+n_\mu n_\nu$, $ ^{(4)}\tilde g_{\mu\nu}=\bar\omega^2 {^{(4)}}\bar g_{\mu\nu}$, with $n_\mu$ the normal to the brane and $\omega$ the dilaton field, then ${^{(4)}}\bar g_{\mu\nu}$ is conformally flat.
It is the contribution from the bulk which determines the real pole on the effective four-dimensional spacetime.
There is no objection  applying 't Hooft's back reaction method in constructing the unitary S-matrix for the Hawking radiation.
Again, there is no "inside" of the black hole.

\end{abstract}
\vspace{2pc}
\section{Introduction}\label{sec:intro}
It is commonly believed that a star with enough mass can collapse into a black hole (BH).
General Relativity Theory (GRT) describes this end-stage very well. There is only a small number of realistic solutions in GRT. Examples are the Schwarzschild and Kerr BH. Astronomers conjecture that almost every galaxy will holds a spinning Kerr BH in its center.
BH's are characterized by  singularities screened to outside observers by an event horizon.
In was realized that the laws of thermodynamics can be applied, resulting in the no-hair theorem, i.e., the only externally observable classical parameters are the  mass, electric charge, and angular momentum.
Hawking, however, realized that a BH will emit thermal radiation\cite{Hawking1975}, which is in a mixed state quantum mechanically. 
However, if a BH is formed out of a pure state and will eventually evaporate into a mixed state, one encounters a serious problem, i.e., the violation of unitarity of quantum mechanics (QM).
So there is a conflict, which is characteristic of the sought-after quantum-gravity model.
One could say that  BH's are the most suitable playgrounds for probing a theory of quantum gravity.
The last decades, several attempts were made to overcome the problems emerging from quantum-gravity,
notably, the information-, the complementarity- and the firewall problems. 
Because the information cannot stay inside the BH after the evaporation, it must in some way come out with the Hawking radiation. 
Further, in order to restore unitarity,  this would entail that the Hawking radiation is not truly thermal.
Many attempts were made solving these paradoxes. The most interesting one is formulated by the complementarity
between the inside and outside of the BH. The information is simultaneously reflected and passed through the horizon.
But then a new problem arises. The particles will be entangled. An outgoing particle at time $t$ must be entangled with all previously emitted radiation\cite{suss1993,page1993}.
It was inescapable to introduce then a firewall of extreme energy close to the horizon\cite{alm2013}. However, an in falling observer will not notice the firewall at all due to the equivalence principle. They will perceive spacetime as Minkowski. Another possibility would be that the in-falling observer is "burned" at the firewall\cite{pol2017}.
In all these models, one ignores, however, the gravitational interaction in a dynamical setting. 
The energy of particles in the vicinity of the horizon will grow beyond the Planck mass.
Also, gravitational waves will come into play and the surface-gravity must be investigated\cite{barrabes2013,slagter2021}.

We will focus in this manuscript on the method developed by 't Hooft\cite{san1987,fol1987,thooft2015,thooft2016,thooft2018,thooft2018a,thooft2018b,thooft2018c,thooft2019,
thooft2021,groen2020, bet2016}.
It is proposed that by a cut-and-past method one replaces the "hard" particles by "soft" particles. This solves the firewall paradox as well as the information paradox by using GRT only (Shapiro effect).
However, one needs in this approach a change in the topology (or  boundary condition). It is called the antipodal mapping. The spacetime inside the BH is removed and the edges are glued together by identifying antipodal points. This idea originates from Schr\"odinger's "elliptic" interpretation\cite{schrod1957}. In the Penrose diagram, region II is the antipode of region I. In fact, region II refers to the same black hole. In-going particles crossing the horizon emerge at the opposite hemisphere and  are strongly entangled, implying a deviation from purely thermal behavior.  So nothing will escape the interior of the BH, because there is no interior.
This also means that time inversion will take place and creation and annihilation operators are interchanged.
An important conclusion is now that the Hartle-Hawking vacuum remains a pure state in stead of a thermodynamically mixed state.
The hypersurface of the horizon can be presented as a M\"obius strip, which is the orientation-reversing stereographic projection of the Riemann sphere. 
We will extend this method to a dynamical axially symmetric five dimensional warped black hole spacetime in conformal dilaton gravity (CDG)\cite{thooft2015b,codello2013,alvarez2014,slagter2019c,slagter2021a}.
CDG fits very well the antipodal identification, because this transformation is part of the conformal group of transformations\cite{felsager1998}.

Conformal invariance can be an exact symmetry, which is spontaneously broken, comparable with the  Brout-Englert-Higgs (BEH) mechanism in the Standard Model (SM) of particle physics.
The spacetime can conformally mapped by $g_{\mu\nu}= \Omega^{4/(n-2)} \tilde g_{\mu\nu}$.
The conformal invariance will be broken by a mass term in the Lagrangian.
The conformal factor $\Omega$ will be used in the complementarity issue
and is probably related to the upper limit of the amount of information that can be stored at the horizon.
To that end, one isolates from the metric the scale dependency, which is called a dilaton field, to be treated on equal footing as a scalar field. One treats this dilaton as an independent dynamical freedom in stead of evaluating the path integral as a perturbative series in the metric components. 

The introduction of a warped five-dimensional spacetime seems rather natural\cite{slagter2022a}. 
The isolation of the dilaton field can be compared with the "warpfactor" in the Randall Sundrum (RS) five dimensional brane world models\cite{randall1999a,randall1999b,ark1998,shirom2000,shirom2003}.
The RS model  was also applied on a Friedmann-Lema\^itre-Robertson-Walker (FLRW) spacetime\cite{slagterpan2016}, where an exact solution of the warpfactor was found. 
The Einstein equations on the brane will be modified by the very embedding in the "bulk". The contribution of the projected Weyl tensor carries information of the gravitational field  outside the brane.
So the brane world observer can be subject to influences from the bulk.
The SM fields are confined to the four-dimensional brane, while gravity acts also in the fifth dimension. 
The model possesses $\mathds{Z}_2$-symmetry, which means that when one approaches the brane from one side and go through it, one emerges into the bulk that looks the same, but with the normal reversed.
A pleasant side effect would be the solution of the hierarchy problem and dark matter issue\cite{maartens2010}. 
During the evaporation process, the warp factor will gradually transform into the dilaton field. Different observers will then use the same $\tilde g_{\mu\nu}$ but different $\omega$

There are some remarkable features of the BH solution\cite{slagter2022a}, which need a follow-up research.
First, the  conformal Laplacian follows from the Einstein equations, so the  dilaton equation turns out to be superfluous.
Secondly, the description of the  antipodal boundary condition by means of the M\"obius strip in the 4D model, can be extended by considering the  Klein surface, which can be embedded in $\mathds{R}^4$. 
Thirdly, we can apply the  Hopf mapping $S^3\rightarrow \mathds{C}\times\mathds{C}$, which is possible in a 5D manifold.
Finally, the quintic solution conjectures a deep seated connection with the symmetries of the icosahedron group and the embedding of the Klein surface in $\mathds{R}^4$.
In this manuscript, we will investigate these relations.

In section 2 we summarize the exact solution. In section 3 we analyze the singularities in the complex plane in relation with the icosahedron group.
In section 4 we revisited the antipodicity and Klein surface   and in section 5 we made some remarks on the Hawing radiation.
\section{Summary of the black hole solution in conformal dilaton gravity }
\label{sec2}
\subsection{The solution\label{sec2-2}}
In a former study\cite{slagter2021,slagter2022a} we found an dynamical exact black hole solution solution on a five-dimensional warped spacetime in conformal dilaton gravity.
We considered the spacetime
\begin{equation}
ds^2=\omega(t,r,y)^{4/3}\Bigl[-N(t,r)^2dt^2+\frac{1}{N(t,r)^2}dr^2+dz^2+r^2(d\varphi+N^\varphi(t,r)dt)^2+d\mathpzc{y}^2\Bigr],\label{2-1}
\end{equation}
where $\mathpzc{y}$ is the extra dimension  and $\omega$ a warp factor in the formulation of RS 5D warped spacetime with one large extra dimension and negative  bulk tension $\Lambda_5$. $\omega$ can also be interpreted as a  dilaton field in conformal gravity models. 
The considered Lagrangian is
\begin{eqnarray}
S=\int d^nx\sqrt{- g}\Bigl[\frac{1}{2}\xi \omega^2 R+\frac{1}{2} g^{\mu\nu} \partial_\mu\omega\partial_\nu\omega+\Lambda\kappa^{\frac{4}{n-2}}\xi^{\frac{n}{n-2}}\omega^{\frac{2n}{n-2}} \Bigr],\label{2-2}
\end{eqnarray}
which means that it is  invariant under
\begin{equation}
g_{\mu\nu}\rightarrow \Omega^{\frac{4}{n-2}} g_{\mu\nu},\quad \omega \rightarrow \Omega^{-\frac{n-2}{2}}\omega.\label{2-3}
\end{equation}
In our  five-dimensional case, we write 
\begin{equation}
^{(5)}{g_{\mu\nu}}=\omega^{4/3} {^{(5)}{\tilde g_{\mu\nu}}},\label{2-4}
\end{equation}
with $\tilde g_{\mu\nu}$ the "un-physical" spacetime. Moreover, in the RS model one writes ${^{(5)}\tilde g_{\mu\nu}}={^{(4)}\tilde g_{\mu\nu}}+n_\mu n_\nu$ with $n^\mu$  the unit normal to the brane. Further, we wrote ${^{(4)}}{\tilde g_{\mu\nu}}=\bar\omega^2 \bar g_{\mu\nu}$.

The exact solution can be represented as
\begin{eqnarray}
\hspace{-1.5cm}\omega=\Bigl(\frac{a_1}{(r+a_2)t+a_3r+a_2a_3}\Bigr)^{\frac{1}{2}n-1}, \cr 
\hspace{-1.5cm}N^2\equiv\frac{N_1(r)}{N_2(t)}=\frac{1}{5r^2}\frac{10a_2^3r^2+20a_2^2r^3+15a_2r^4+4r^5+C_1}{C_2(a_3+t)^4+C_3}=\frac{4\int r(r+a_2)^3dr}{r^2[C_2(a_3+t)^4+C_3]},\cr
\hspace{-1.5cm}N^{\varphi}=F_n(t)+\int\frac{1}{r^3\omega^{\frac{n-1}{n-3}}}dr,\label{2-5}
\end{eqnarray}
with $a_i$, $C_i$ some constants and $n=4, 5$. Further we took $a_3=a_2a_3$. $F_n$ is a arbitrarily function in t,  determined by constraint equations.
The two solutions for $\omega$ and $\bar\omega$ are represented  by $n=5$ and $n=4$ respectively. The equation for $N$ is the same, as it should be, apart from the constants.
The solution shows that the the 5D Einstein equations and the effective 4D Einstein equations must be solved together. They don't form separately  a closed system\cite{shirom2003}. One can easily check that the Einstein equations are traceless and that the conservation equations are fulfilled. 

The polynomial $N$ is a quintic. This is solely caused by the contribution from the bulk, i.e., the 5D Weyl tensor which carries information of the gravitational field outside the brane.
Without the 5D contribution on the effective 4D Einstein equations, one obtains the solution for $N^2$ and $\omega$
\begin{equation}
N^2=\frac{1}{4r^2}\frac{(6d_2^2r^2+8d_2r^3+3r^4+D_1)}{D_2(t+d_3)^2+D_3},\quad \omega=\frac{1}{(r+d_2)t+rd_3+d_2d_3}.\label{2-6}
\end{equation}
Again, one can write the r-dependent part as (compere with eq.(\ref{2-5}))
\begin{equation}
N_1(r)^2=\frac{3}{r^2}\int r(r+d_2)^2dr.\label{2-7}
\end{equation}

The solution eq.(\ref{2-6}) can be compared with the Ba\u nados-Teitelboim-Zanelli (BTZ) black hole\cite{banadoz1992} in $(2+1)$-dimensional spacetime
\begin{eqnarray}
N^2=\frac{\frac{r^4}{l^2}-8MGr^2+16G^2J^2}{r^2}=\frac{4}{l^2r^2}\int r(r+ l\sqrt{4GM})(r- l\sqrt{4GM}),\cr N^\varphi=-\frac{4GJ}{r^2},\label{2-8}
\end{eqnarray}
with $M$ the mass parameter and $J$ the angular momentum\cite{compere2018}. It is evident that our constants $a_i$ must be related to the mass.

$(2+1)$-dimensional gravity has been recognized as a laboratory  for studying GRT in connection with quantum-gravity issues.
When one omits the $dz^2$ and $d\mathpzc{y}^2$ terms in eq.(\ref{2-1}), one can solve the Einstein field equations with on the right hand side only a cosmological constant $\Lambda$. For negative value of $\Lambda=-\frac{1}{l^2}$, the solution is locally  anti-de Sitter.
The length scale $l$ determines the distance at which curvature sets in.
This solution is non trivial. There are topological properties, such as local defects and black holes.
It is not a surprise that these models are used in constructing quantum gravity models, because locally one deals with Minkowski  spacetime. So  planar gravity fits in very well.

This BTZ solution shares some features with the 4D Kerr solution. The equation for $N^2$ in eq.(\ref{2-6}) has two positive roots. One of  these represents a Killing horizon.
The relation with the dynamical "uplifted"" BTZ solution was presented by Slagter\cite{slagter2019b}. In this case the $dz^2$ term is maintained. It is remarkable that $\Lambda$ must then be taken zero. 
After the discovery of the  AdS/CFT correspondence, the BTZ solution gained new interest and became a tool to understand black hole entropy. It is not yet clear whether pure 3D Einstein gravity make sense quantum mechanically without string theory embedding.
\section{Treatment of the singularities and complex analysis\label{sec:3}}
\subsection{The quintic in the complex plane}\label{sec:3-1}
The singularities are determined by a quintic polynomial in $r$ (see eq.(\ref{2-5}); we renamed the constants). This polynomial, when equated to zero, can have complex solutions. See figure 1.
\begin{figure}[h]
\centering
\resizebox{0.4\textwidth}{!}
{\fbox{\includegraphics{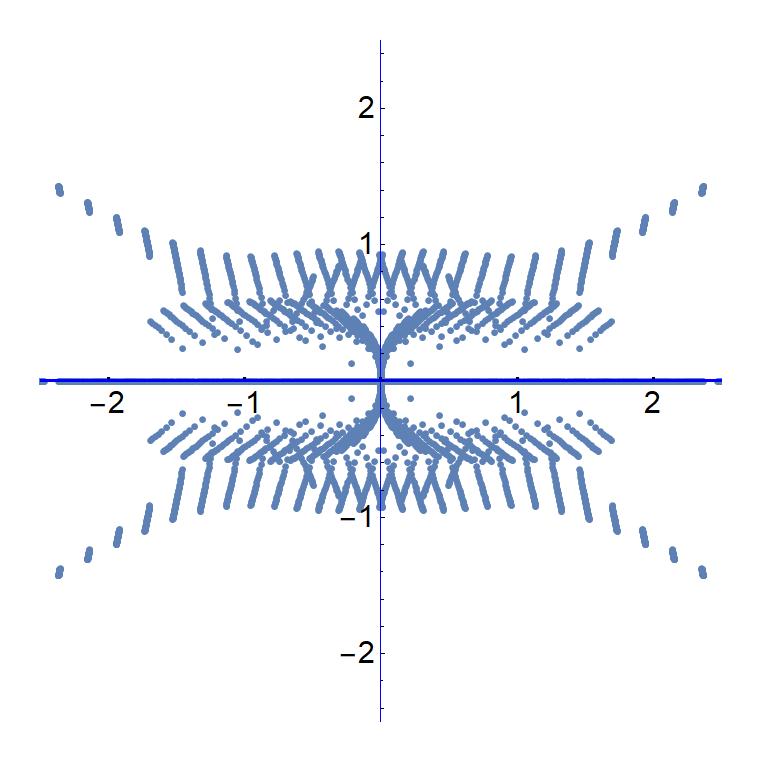}}}
\caption{ Location of the roots of $\frac{1}{5r^2}\Bigl(4r^5-15ar^4+20a^2r^3-10a^3r^2+c\Bigr)$ in the complex plane by varying the constants $-4<a<4, -4<c<4$.}
\label{fig:1}       
\end{figure}
Singularities in complex analysis come in different levels of "badness". We will investigate these singularities on the Riemann sphere. The stereographic projection $\pi: S^2\setminus \{N\}\rightarrow\mathds{R}^2$ is conformal and orientation reversing, needed for the antipodicity. See figure 2
\begin{figure}[h]
\centering
\resizebox{1.1\textwidth}{!}
{\includegraphics[width=1.6cm]{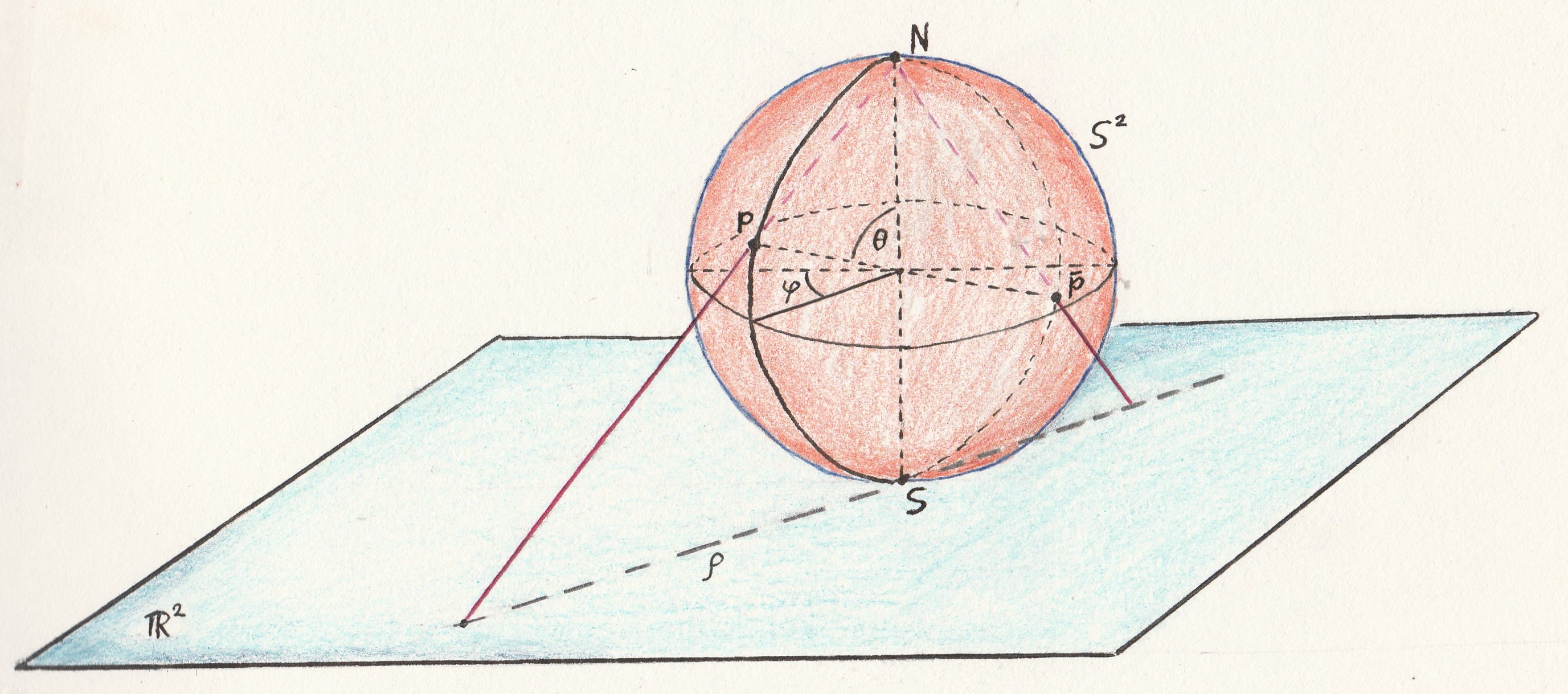}\hspace{-8pt}
\includegraphics[width=1.1cm]{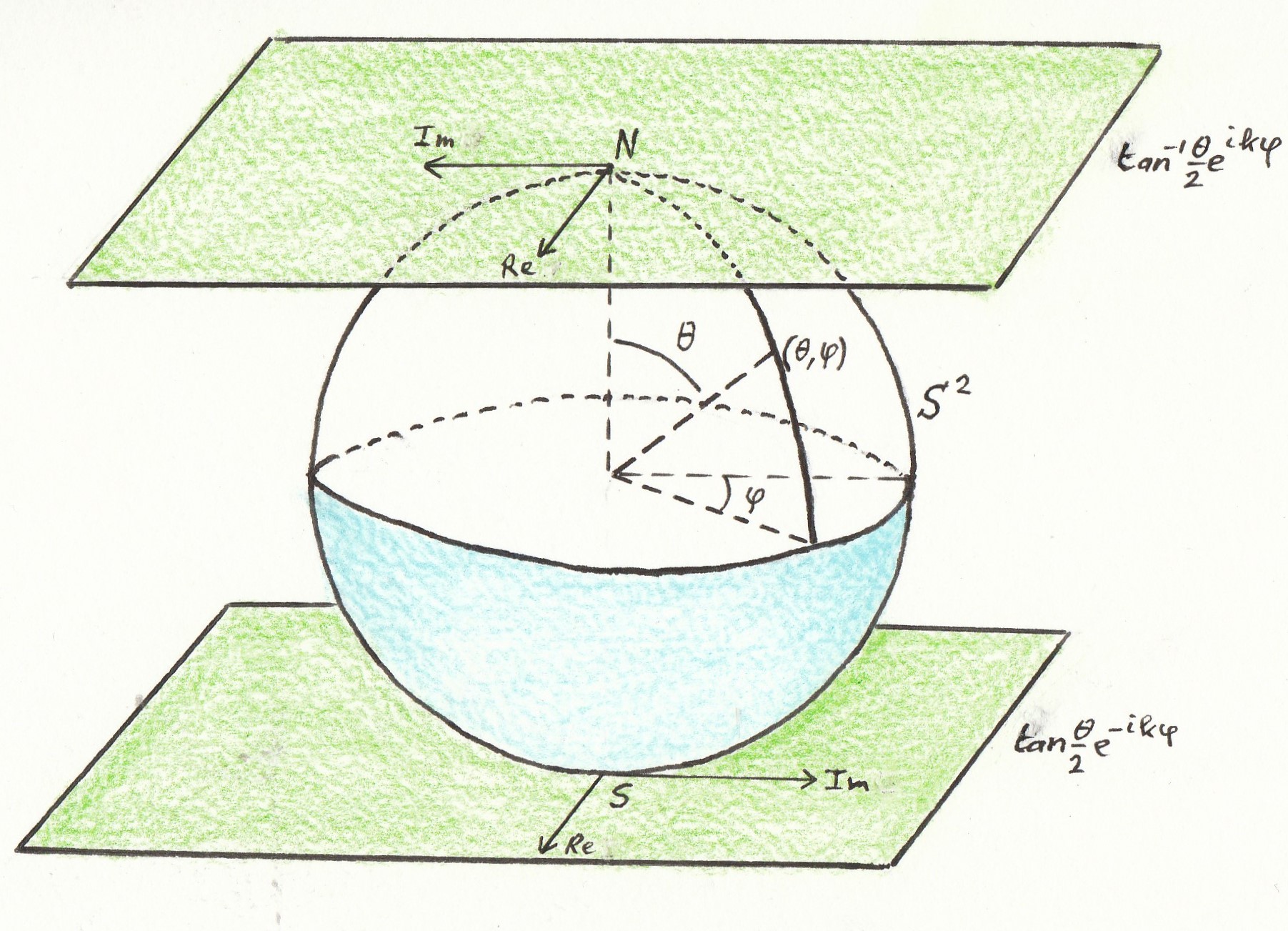}}
\caption{Left: stereographic projection $\pi: S^2\setminus \{N\}\rightarrow\mathds{R}^2$, which is a conformal and orientation reversing map. By adding a point at infinity, one compactifies the plane, because the inversion (reflection in the equator) is otherwise not defined. The $\mathds{R}^2$ and $S^2\setminus\{N\}$ are now neatly conformal. Right: the Riemann sphere, a complex manifold with the map $S^2\rightarrow \mathds{C}$.}
\label{fig:2}  
\end{figure}
The singularities  can be  removable, a  pole or an  essential singularity. For example, for $c=0$, the singularity $r=0$ is trivially removable.
However, we want to investigate, for $c\neq 0$, if our singularities are essential on the Riemann sphere.
A fundamental theorem of the algebra says that every complex polynomial in n will have a zero. We will consider $r$ now complex. So we shall temporarily replace $r$ by  ${\bf z}$, not to confuse with the Cartesian $z$. We will write later on $r=Re^{ im\varphi}$. 
We know that a streographically  polynomial map $K$ from the projective plane to itself, corresponds to a map $f$ from the sphere to itself.
Then $ f=\pi_N^{-1} K \pi_N$, with $\pi_N$ the stereographic projection of the Riemann sphere to the plane. This map is smooth everywhere, even in the neighborhood of the north pole.
We set $L=\pi_S f\pi_S^{-1}$. One then proves, by using $\pi_S\pi_S^{-1}=\frac{1}{{\bf \bar z}}$ and  $K(z)={\bf z}^n+a_1{\bf z}^{(n-1)}+... + a_n$ , that
\begin{equation}
L({\bf z})=\frac{{\bf z}^m}{K({\bf z})},\quad m\leq n\label{3-1}
\end{equation}
is also smooth in the neighborhood of 0. Further  $f=\pi_S^{-1}L\pi_S$ is smooth in the neighborhood of N.
In our case we are dealing  with a ratio 
\begin{equation}
F(r)\equiv \frac{P(r)}{Q(r)}=\frac{4r^5-15ar^4+20a^2r^3-10a^3r^2+c}{5r^2}=\frac{\prod\limits_{i=1}^{5}\alpha_i(r-r_i)}{5r^2},\label{3-2}
\end{equation}
with $\alpha_i$ are constants.
The inverse\cite{slagter2021} 
\begin{equation}
\frac{1}{F(r)}=\frac{Q(r)}{P(r)}=\frac{d}{dr}\Bigl[\sum_{r_i}\frac{r_i\log_c(r-r_i)}{4(r_i-a)^3}\Bigr],\label{3-3}
\end{equation}
determines the singularities of our spacetime. Here the sum is over the roots of the quintic.
The subscript c stands for the complex logarithm. $F(r)$ is holomorphic outside the zero of $Q(r)$. Note that some roots can have multiplicities. We can apply now eq.(\ref{3-1}).
$f$ has only a finite number of critical points, because $P'=20{\bf z}({\bf z}-a)^3$ is not identically zero.
The set of regular values of $f$ with a finite number of points removed from the sphere, is therefore connected.
Remember that a holomorphic map from the Riemann sphere into itself, $f: S^2\rightarrow S^2$,  can be presented, by using standard coordinates via the stereographic projection, as a ratio of polynomials $\frac{P({\bf z})}{Q({\bf z})}$.  An orientation-preserving (or reversing) map $f$ is conformal if $F({\bf z})$ can be written as an algebraic function
\begin{equation}
F({\bf z})=\frac{P({\bf z})}{Q({\bf z})} \qquad \Bigl(  =\frac{P(\bar {\bf z})}{Q(\bar {\bf z})}\Bigr).\label{3-4}
\end{equation}
One also proofs that the singularities are poles by using the properties of the Riemann sphere and the multiple valuedness of the  transcendental complex logarithm, $\log_cr=\log|r|+i \arg{r}+2k\pi i$ in eq.(\ref{3-3}). 
However, the derivative does not depends on the branch $k$. So we obtain
\begin{equation}
\frac{1}{F}=\frac{Q}{P}=\sum_{r_i}\frac{r_i}{4(r-r_i)(r_i-a)^3}.\label{3-5}
\end{equation}
We can use the properties of Laurent series (and  Weierstrass theorem) and the Riemann sphere, to see that we have no essential singularities.
In figure 3 we plotted the logarithm of the polynomial $F$.
\begin{figure}[h]
\centering
\resizebox{0.8\textwidth}{!}
{ \includegraphics{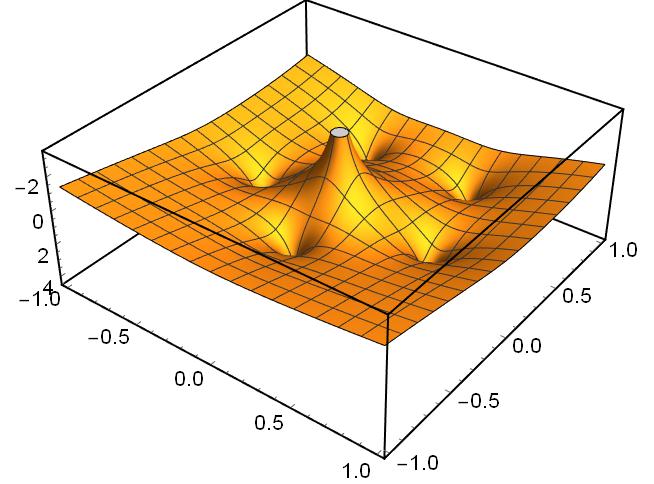}\includegraphics{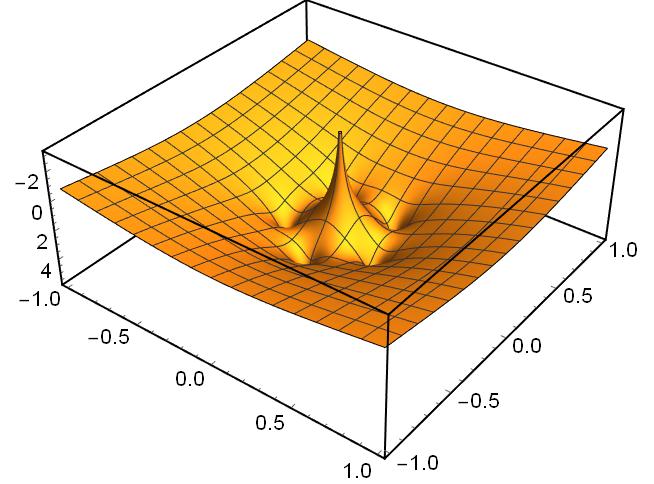}}
\caption{ Left: location of the roots of $F(r)$ in the complex plane for the real values  $a= c=0.3$. One plots the logarithm of the absolute value of the polynomial. There is only an essential singularity at ${\bf z}=0$. Right: the same plot for a smaller value of $a$. Note the striking  similarity with figure 10 of the appendix A of stereographically projected vertices of the icosahedron, put in a position such that the line north-south is through two vertices.}
\label{fig:3}       
\end{figure}
\begin{figure}[h]
\centering
\resizebox{1\textwidth}{!}
{ \includegraphics{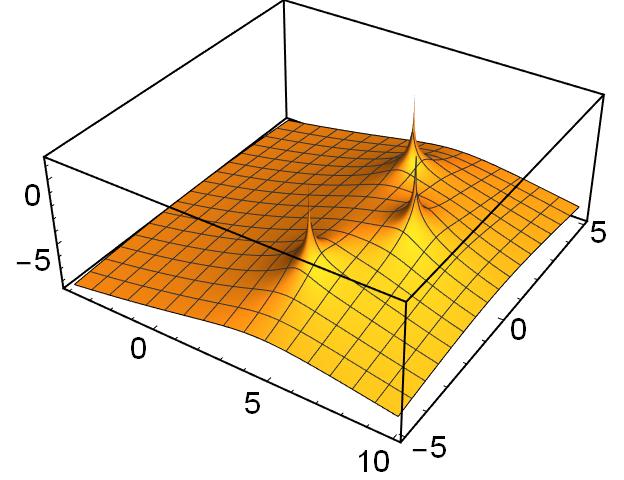}\includegraphics{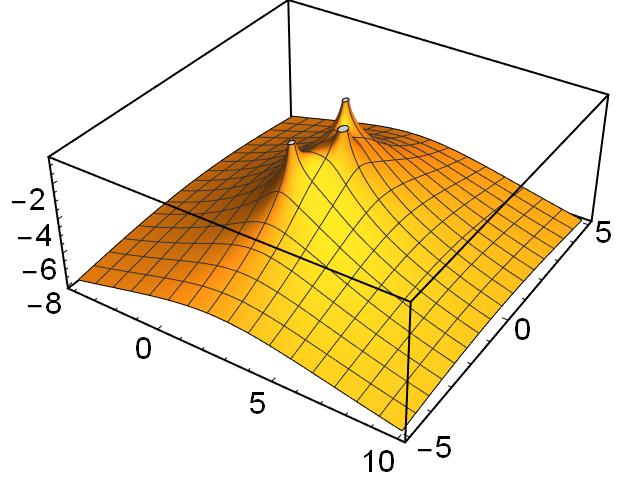}\includegraphics{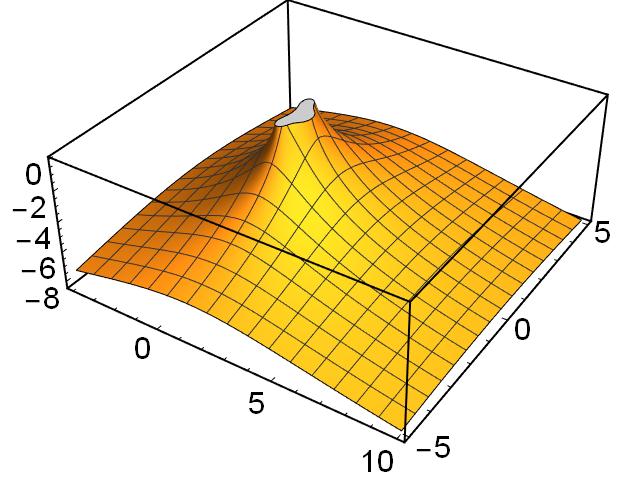}\includegraphics{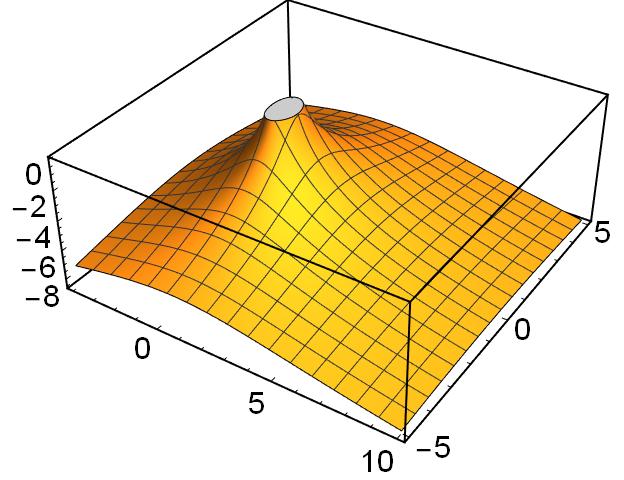}}
\caption{ Location of the roots of the residue $R$ in eq.(\ref{3-6}) in the complex plane for several real values of a. From left to right, 4, 2, 1 and 0.5. For negative values, the behavior is almost the same.}
\end{figure}
One can isolate the $r=0$ pole by writing $F$ as
\begin{eqnarray}
F=-\frac{1}{5}r^3+r^2(r-a)-2r(r-a)^2+2(r-a)^3+\frac{c}{5r^2}=\frac{c}{r^2}+R(r).\label{3-6}
\end{eqnarray}
We can plot $R(r)$ for several values of $a$. See figure 4.
The Laurent polynomial has  a pole at $r=0$. When $a$ decreases to small values, we observe that there is 1 pole at $r=0$.
Often, the ratio of two holomorphic polynomials is called a  meromorphic function. Their poles are isolated. Our $\frac{1}{N_1^2}$ is meromorphic ($Deg(P)>Deg(Q)$), which determines the singular points of the black hole spacetime. 
Further, $S^2$ is isomorphic to $\mathds{C}_\infty$, and $\mathds{C}_\infty\rightarrow \mathds{C}_\infty$ is holomorphic.
Let $F_o({\bf z})$ be a regular smooth map. The preimage consists then of the solutions of
\begin{equation}
P({\bf z})-F_oQ({\bf z})=0.\label{3-7}
\end{equation}
In our case, $Deg(P)=5$ and so  we expect 5  distinct solutions (with multiplicities). Further, ${\mit\Psi}^a$ has winding number $n=-5$.
One calls eq.(\ref{3-7}) the  polyhedral equation associated with the finite M\"obius group G with degree $n$. See next sections.
There are 3 distinct cases of interest: $F_o=(0, 1, \infty)$, corresponding with the singularities, the regular case and the pole $r=0$ respectively. Eq.(\ref{3-7}) represents  a  Laurent series. 
Note that for $c= a^5$,  we have the special case $F=\frac{(r+\frac{a}{4})(r-a)^4}{5r^2}$
We can now write $r$  as  
\begin{equation}
r=Re^{in\varphi}.\label{3-8}
\end{equation}
The distinct solutions can then be characterized by only one pole $R=0$, and $n$ rotations in the complex plane.
This was conjectured: the fractional M\"obius transformations of the icosahedron group, as we shall see in the next sections.
\subsection{The icosahedron group, elliptic functions and the quintic}\label{sec3.2}
Because the exact solution of our conformal black hole  on the 5D warped spacetime is determined by a quintic, we expect that there is a link with the icosahedron group of symmetries. It was Klein who already noticed this correspondence in 1888 \cite{klein1888}. 
There will be a direct relation between the zero's of our quintic and the icosahedron equation.
It is worth noting that the direct isometries of $\mathds{R}^3$  for the icosahedron is the alternating group ${\cal A}_5$, i.e., the 120 elements of the permutation group of 5 entities. 
In the Appendix A we summarize the main features of the icosahedron group and the relation with the M\"obius transformations. 

Suppose, we have a homogeneous polynomial ${\mit \Xi}: \mathds{C}^2\rightarrow \mathds{C}$.  One calls this a {\it form} if ${\mit \Xi}(\lambda {\bf z},\lambda {\bf w})=\lambda^p {\mit \Xi}({\bf z},{\bf w})$, with $\lambda, {\bf z}, {\bf w}\in \mathds{C}$.
Consider now two forms $({\mit \Xi}, {\mit \Pi})$, $G^*$-invariant.
Now we define the map $q_G:\mathds{C}P^1\rightarrow \mathds{C}P^1$ by ($G^*$-invariant)
\begin{equation}
q([{\bf z}:{\bf w}])=[{\mit \Xi}({\bf z},{\bf w}):{\mit \Pi}({\bf z},{\ bf w})],\qquad {\bf z,w}\in \mathds{C}.\label{3-9}
\end{equation}
By the identification  $\mathds{C}P^1=\mathds{C}_\infty$ it becomes a holomorphic map $\mathds{C}_\infty\rightarrow\mathds{C}_\infty$ with rational restriction to $\mathds{C}$:
\begin{equation}
q(\zeta)=\frac{{\mit \Xi}({\bf z,w})}{{\mit \Pi}({\bf z,w})}=\frac{{\mit \Xi}(\zeta ,1)}{{\mit \Pi}(\zeta ,1)}.\label{3-10}
\end{equation}
In order to obtain the invariant forms of the spherical Platonic tessellations, one uses the $C_d=\{\zeta\rightarrow e^{2\pi i\frac{l}{d}}\}$-invariant in order to construct $G^*$-invariance.
For the icosahedron ($G^*=I^*$), the invariant form becomes \cite{Toth2002}
\begin{equation}
1728{\cal I}^5-{\cal J}^2-{\cal H}^3=0,\label{3-11}
\end{equation}
with ${\cal I}$ the Hessian, ${\cal H}=\frac{1}{124}{\cal I}$ and ${\cal J}$ the Jocabian $Jac({\cal I},{\cal H})$. So an $I^*$-invariant form can then be written as a polynomial in the basic invariants $({\cal I}, {\cal J}, {\cal H})$.

For details, we refer to Toth's book\cite{Toth2002}.
We will not go further into details concerning this issue. For more details, see for example  Shurman \cite{Shurman1997}. 
\subsection{Invariant forms and the quintic polynomial\label{sec:3.3}}
In order to find an exact solution of the zero's of eq.(\ref{3-2}), 
one can apply the  Tschirnhauser transformation \cite{slagter2022a}. One obtains
\begin{equation}
F=r^5+\frac{3\cdot 5}{2^4}a(c-a^5)r^2+\frac{5^3}{2^8}a^3(c-a^5)r-\frac{1}{2^4}(c-a^5)^2.\label{3-12}
\end{equation}
This is the  principal form.
The discriminants of the Tschirnhauser and the original form of eq(\ref{3-2}) are different, i.e., 
\begin{equation}
-\frac{5^4}{2^8}c(a^5-c)^3, \qquad -\frac{5^4}{2^{32}}(a^5-c)^5c(3\cdot 7^2\cdot 23a^5-2^8c)^2\label{3-13}
\end{equation}
respectively, which means that there is another special solution for $c=\frac{3381}{256}a^5$ (besides $c=0$ and $c=a^5$). It delivers four complex  and one real solution,
\begin{equation}
r=\frac{a}{4}\Bigl(4-\frac{5}{6}\sqrt[3]{108+12\sqrt{69}}-\frac{10}{\sqrt[3]{108+12\sqrt{69}}}\Bigr).\label{3-14}
\end{equation}
For the three special cases $\sum_i r_i=\frac{15}{4}a$.
An advantage is that the Tschirnhaus form can be compared with the icosahedron equation, by finding invariant  forms.
In general one can state that, in the icosahedron case, the problem is not solvable by radicals only.
One needs the Brioschi reduction
\begin{equation}
r^5-10\alpha r^3+45\alpha^2 r-\alpha^2.\label{3-15}
\end{equation}
The discriminant is $5^5\alpha^8(1728\alpha-1)^2$.
$\alpha$ is given by $\alpha=\frac{1}{1728-J}$, with
{\footnotesize
\begin{eqnarray}
\hspace{-1cm}J=a^{-15}\Bigl[2602551455a^{20}-22605030559a^{15}c+23403011072a^{10}c^2 \hspace{5cm}\cr
\hspace{-1cm}+4045149a^{10} Z -35515526912a^5c^3 -5806848Za^5c+150994944c^4+589824Zc^2\Bigr]^3{\bf \div}\hspace{2cm}\cr
\hspace{-2.4cm}\Bigl[\Bigl(44933a^{15}-1833349ca^{10}+1985024a^5c^2+a^5 143Z-196608c^3-768cZ\bigr)(a^5-c)(17161a^5-1536c)^5\Bigr]\hspace{0.2cm},\label{3-16}
\end{eqnarray}}
and with 
\begin{equation}
Z=\sqrt{-c(a^5-c)(3381a^5-256c)^2}\label{3-17}.
\end{equation}

In order to express the single parameter $\alpha$ in terms of $a$ and $c$, one used the special symmetry properties of the icosahedron, i.e., the invariant polynomials for the icosahedron group acting on the Riemann sphere.
The five roots of the Brioschi quintic are the vertex polynomials of the five octahedra in an icosahedron set. The single parameter $\alpha$ is then obtained from the polyhedral polynomials of the underlying icosahedron. One uses the stereographic projection of the Riemann sphere on the complex plane\cite{king1992} (see Appendix A).
This quintic can also be solved analytically using elliptic curves. See next section.

\subsection{Ultra radical and elliptic functions\label{sec:3.4}}
It is well known that  quadratic, cubic and quartic equations can be solved by radicals. This is due to the the fact that  transcendental functions can be expressed as 
\begin{figure}[h]
\centering
\resizebox{1.\textwidth}{!}
{\fbox{ \includegraphics{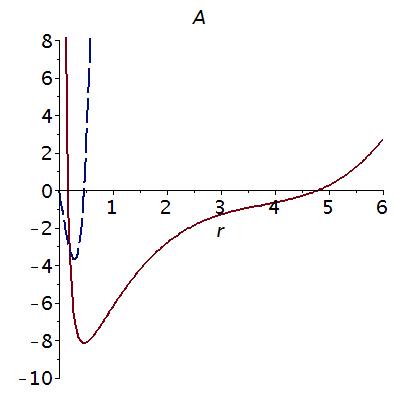}\includegraphics{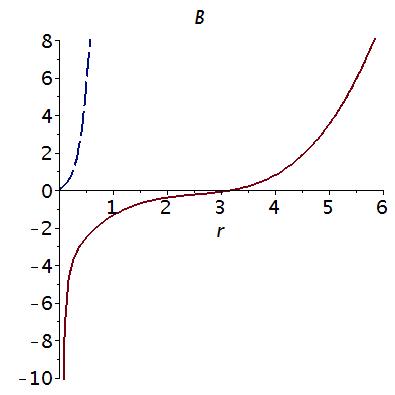}\includegraphics{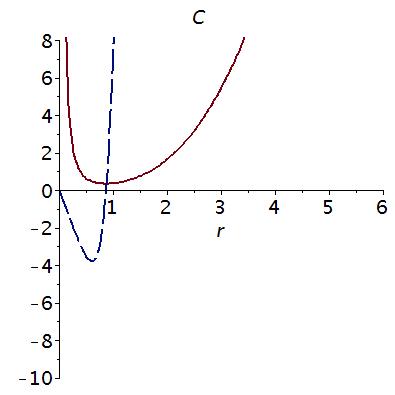}}}
\caption{Plot of the solutions  F(r) for three different values of $a$ and $c$ (line) together with the Wronskian (dashed)}
\label{fig:6}       
\end{figure}
\begin{equation}
f(x)=\int\frac{dx}{\sqrt{\mathpzc{P}(x)}},\label{3-18}
\end{equation}
with  $\mathpzc{P}(x)$ a polynomial. In fact one can use logarithms and trigonometric functions as well. However, not for the quintic\cite{king1996}. This fact finds its origin in the Galois group of the equation. The Galois group of the quintic is the alternating $A_5$, which is a simple group. One needs now  ultra radicals  or  Bring radicals. The Bring form of the quintic is $x^5+d_1x+d_2$ (or $x^5-x+d$, by changing the variable). They turn out to be  elliptic  functions by using eq.(\ref{3-18}). It was Hermite who already recognized in 1858 that elliptic transcendents can play a role in the solution of the Bring quintic . The Bring radical  can be  real valued and  an analytic function in the neighborhood of the real line.
Now $d$ (containing our parameters $a$ and $c$ in eq.(\ref{3-5})) will contain the mass parameter $a$ of the black hole. It is conjectured that in the dynamical process of the black hole evaporation, the behavior of figure 5B will emerge. For decreasing mass, $a$ will tends to zero and the Bring radical will be approached. This means that the zero of our quintic approaches the $r=0$ singularity. Note that the Wronskian has a zero at $r=0$ in the case of figure 5B.
For the  Brioschi form  in eq.(\ref{3-15}) one can take the associated  Weierstrass elliptic  curve
\begin{equation}
\mathpzc{E}_\alpha: y^2+ry=r^3+36\alpha x+\alpha,\label{3-19}
\end{equation}
with $\alpha\neq 0,\frac{1}{1728}$ and discriminant $2^4(1728\alpha-1)^2$.  An elliptic curve defined over a number field $K(a,c)$ has genus one and is a smooth projective curve with a distinguished (rational) point. In our case it must be the real zero. 
One should use a suitable program in order to determine for which  primes and $\alpha\in \mathds{Q}$ (i.e., $a$ and $c$) delivers the single real root\cite{duke2002}, in order to check the solution eq.(\ref{3-16}).
For more details of this peculiar issue, the interested reader should consult the impressive books of King\cite{king1996} and Connell\cite{conn1999}.
\section{Antipodicity and the application of the Klein surface\label{sec:4}}
\subsection{The projective $\mathds{R}P^2$ and $\mathds{R}P^3$\label{sec:4.1}}
Antipodal mappings are tight connected to projective planes $P^n$ in pseudo-Cartesian manifolds.
They are conformal. Let $P^n = \{({\bf x},-{\bf x})\mid{\bf x}\in S^n\}$ and define $\pi: S^n\rightarrow P^n$ by $\pi({\bf x})=\{{\bf x},-{\bf x})$. So $P^n$ consists of all unordered pairs of antipodal points of $S^n$ and $\pi$ takes a point of $S^n$ and pairs it to its antipodal point. So we say that $P^n$ is the identification space obtained by identifying antipodal points of $S^n$, i. e., $\pi({\bf x})=\pi(-{\bf x})$\footnote{We shall see that this identification is just what is needed when considering the horizon in the black hole spacetime.}.
One needs only the upper closed hemisphere and identifies antipodal points on the boundary.
Let us take the disc $B^2$\cite{gauld2006}. To obtain $P^2$ one must identify antipodal points on the boundary, say a, b and c. See figure 6. The two lines connecting a and c, when put together, represents a circle embedded in $P^2$. This circle bounds two regions in $P^2$: a disk obtained by identifying the relevant parts of the boundary. The quadrilateral represents a M\"obius strip.
Thus $P^2$ is obtained by gluing together a M\"obius strip and a disk along their edges.
One calls this a cross cap. So we obtain the $\mathds{R}P^2$ from $S^2$ (see also figure 3). The latter is a realization of $\mathds{R}P^2$, which parametrizes straight lines passing through the origin in $\mathds{R}^3$. 
\begin{figure}[h]
\centering
\resizebox{0.4\textwidth}{!}
{\includegraphics[width=.90cm]{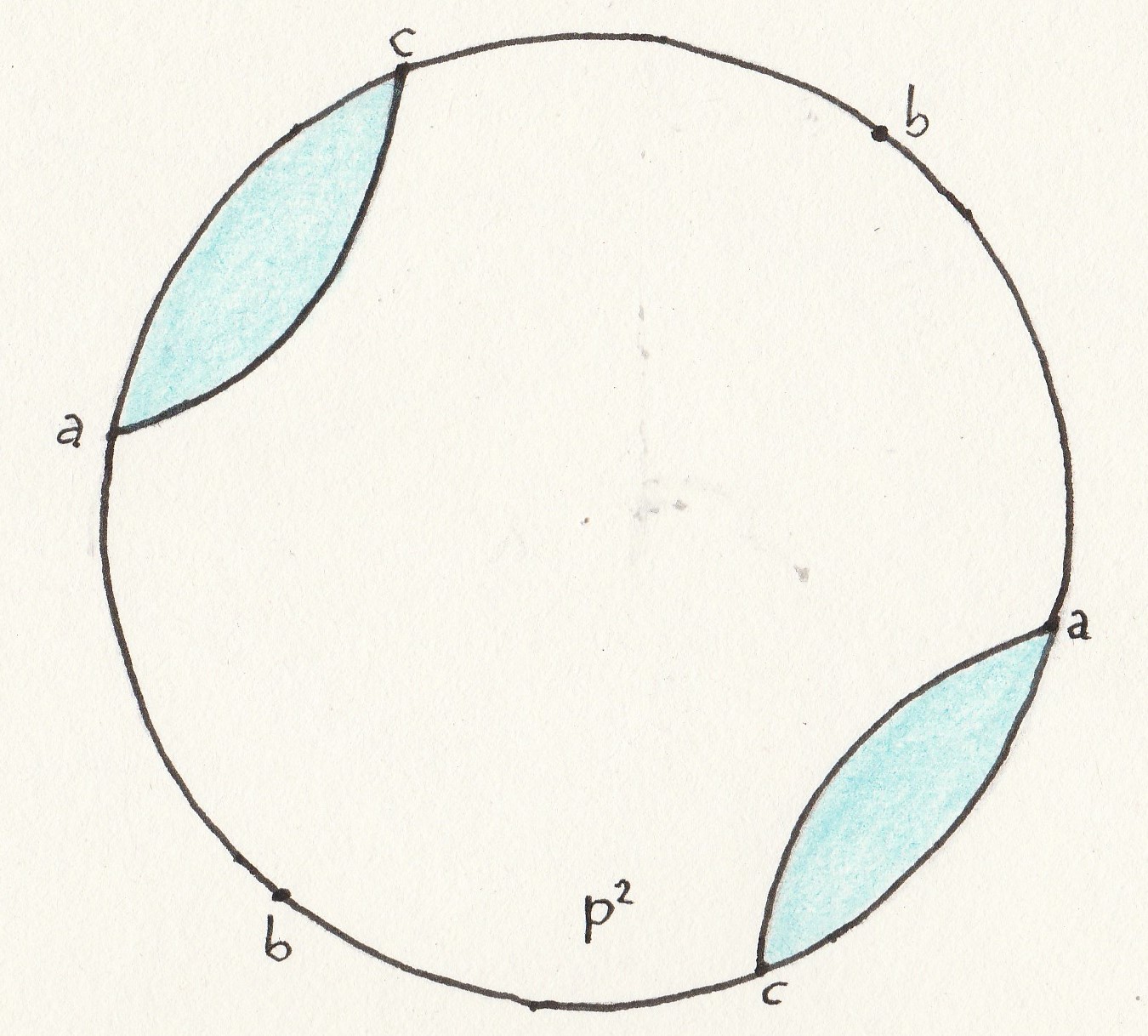}}
\caption{Construction of the  cross cap }
\label{fig:7}
\end{figure}
In the one dimension higher case, we consider the closed ball $B^3(R)$. The antipodal points on the boundary sphere $S^2(R)$ are again identified. 
So a particle that passes the boundary to the inside, re-emerges at the antipodal point on the boundary. 
Now the $\mathds{R}P^3$ parametrizes straight lines through the origin in $\mathds{R}^4$.
\subsection{The complexification of the warped spacetime\label{sec:4.2}}
One can parameterize the Klein bottle on different ways. We are interested in the cases where the projected $S^2$ is our horizon.
Remember that  the stereographic projection on the plane was the M\"obius strip, where antipodal points were identified.
One can parametrize the central line, say $\alpha(\tau)$ and the radius $r(\tau)$ in different ways (see figure 7). The central line is here parametrized as
\begin{equation}
x_1=a(1-\cos\tau),\qquad x_2=b\sin\tau(1-\cos\tau),\qquad R(\tau)=c-d(\tau-\pi)\sqrt{\tau(2\pi-\tau)}\label{4-1}
\end{equation}
for the Cartesian coordinates $(x_1,x_2)$ and where $R$ represents the radius.
\begin{figure}[h]
\centering
\resizebox{0.4\textwidth}{!}
{\includegraphics[width=10cm]{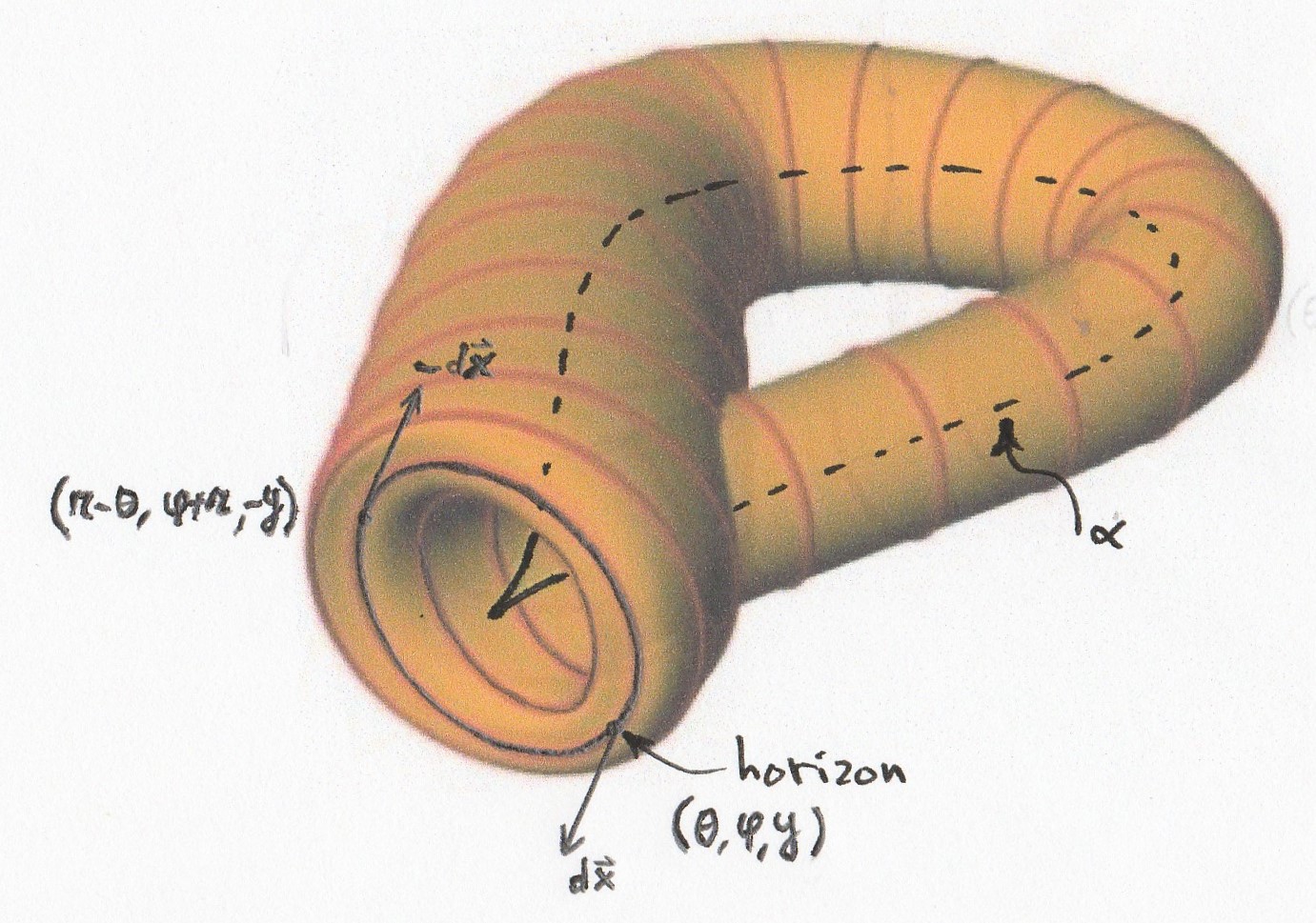}}
\caption{ Plot of the central line $\alpha(\tau)$ ("directrix")  of the Klein surface and the schematic embedding in $\mathds{R}^4$  }
\label{fig:8}       
\end{figure}

One should like to have $\alpha(a)=\alpha(b), \alpha'(a)=\alpha'(b), r(a)=r(b), r'(a)=r'(b)=\pm\infty$, so the two tube ends must meet tangent-wise along the common boundaries in $a$ and $b$. Further, $\parallel\alpha'\parallel$ must be everywhere non zero. 
Now we apply this model to the horizon. When approaching from region I in the Penrose diagram the horizon, one expects to enter  region II. However, one shows up at the antipode, for example in spherical polar coordinates, $(-U,-V,\pi-\theta,\varphi+\pi,-y)$ at "the opposite side" of the black hole (we could equally work in polar coordinates $(z,\varphi,y)$). As already mentioned in the introduction, this has very pleasant consequences concerning the entanglement issues \cite{thooft2018}. 

The Klein bottle can be seen as an union of two M\"obius strips. It is homeomorphic to the union of two copies of a M\"obius strip joined by a homeomorphism along their boundaries. So the Klein bottle is the connected sum  of two projective planes. See figure 8.
\begin{figure}[h]
\centering
\resizebox{0.7\textwidth}{!}
{{\includegraphics[width=.25cm]{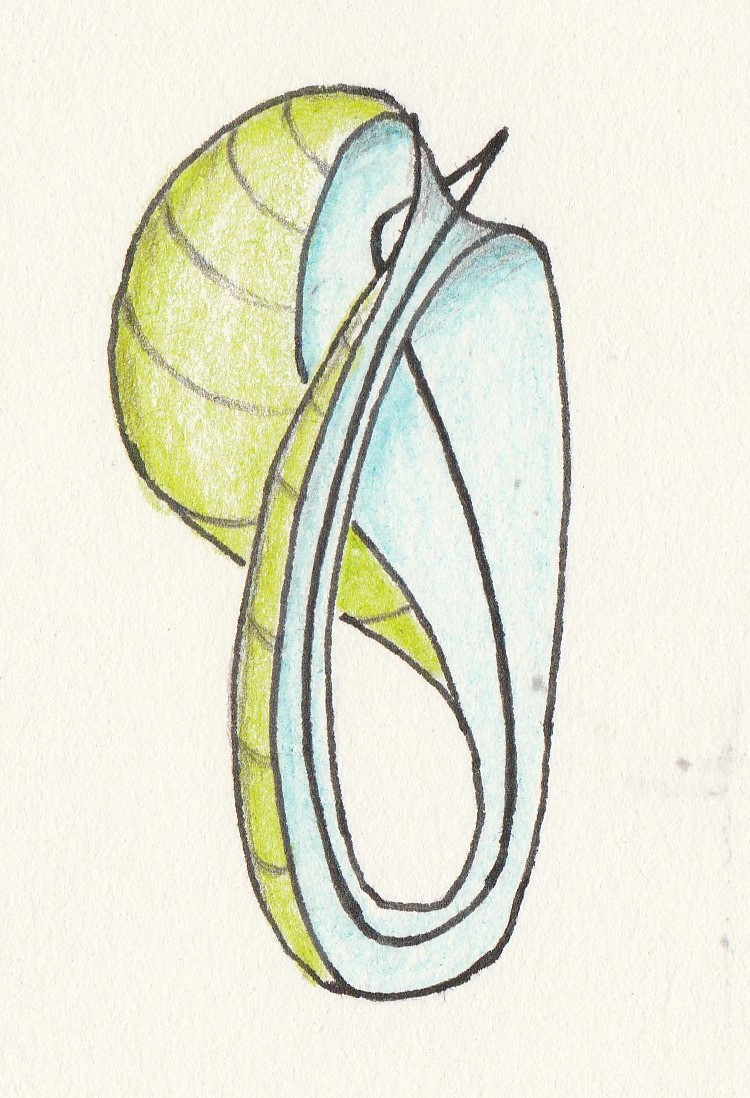}\hspace{-4pt}
\includegraphics[width=.25cm]{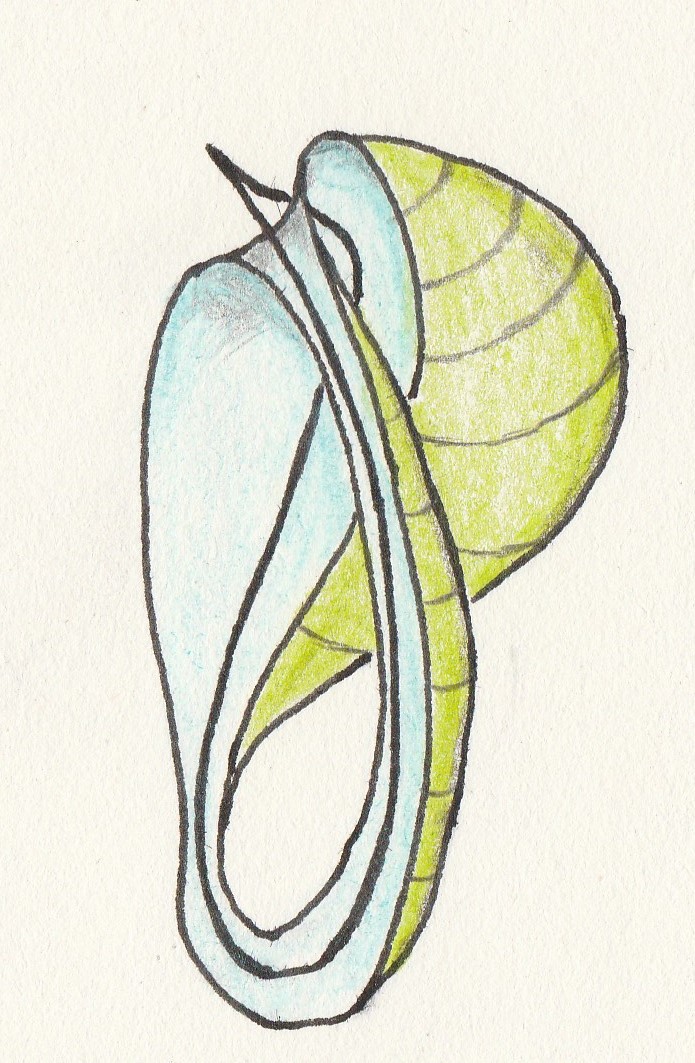}\hspace{-4pt}
\includegraphics[width=.36cm]{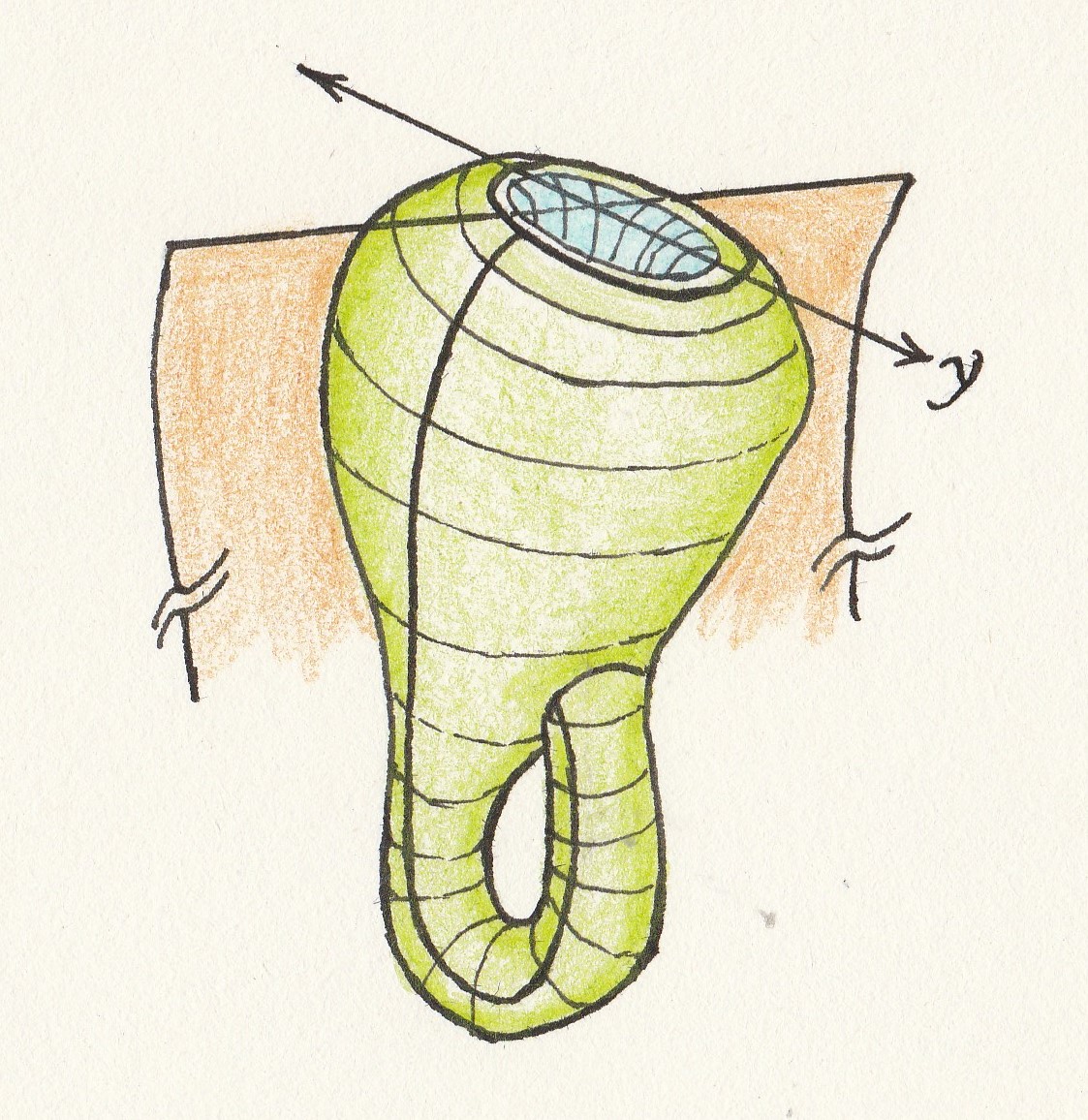}}}
\caption{Klein bottle as two M\"obius strips merged and embedded in $\mathds{R}^4$. $\mathpzc{y}$ represents the bulk space coordinate}
\label{fig:9}
\end{figure}
We already know that $\mathds{R}^4$ is homeomorphic with $S^1\times\mathds{R}^3$. 
If we have $(a,b,c,d)\in \mathds{R}^4$ with $ad>bc$, then we can express the matrix
\begin{equation*}
\begin{pmatrix}
a& b \\
c & d 
\end{pmatrix}\label{4-3}
\end{equation*}
as a product
 \begin{equation*}
\begin{pmatrix}
\cos\varphi& \sin\varphi \\
-\sin\varphi& \cos\varphi
\end{pmatrix}
\begin{pmatrix}
p& q \\
0& r
\end{pmatrix},\label{4-4}
\end{equation*}
with $p,r >0$.

We will now complexify the homeomorphism.
Let us  write our coordinates $( r, z, \mathpzc{y},\varphi^*)$ (with $x=r\sin\varphi, y=r\cos\varphi$)
\begin{equation}
{\cal V}=z+i\mathpzc{ y}=Re^{i\varphi},\qquad {\cal W}=x+iy=re^{i\varphi},\label{4-5}
\end{equation}
where the antipodal map is now ${\cal V}\rightarrow -{\cal V}\equiv-R^2/\bar{\cal  V} , {\cal W}\rightarrow -{\cal W}\equiv-r^2/\bar{\cal  W}$.

Further, ${\cal V}\bar{\cal V}=z^2+\mathpzc{y}^2=R^2,{\cal W}\bar{\cal W}=x^2+y^2=r^2 $.
And after inversion
\begin{eqnarray}
z=\frac{1}{2}({\cal V}+\bar{\cal V}), \qquad \mathpzc{y}=\frac{1}{2i}({\cal V}-\bar{\cal V}),\hspace{0.5cm}\cr
x=\frac{1}{2}({\cal W}+\bar{\cal W}),\qquad y=\frac{1}{2i}(({\cal W}-\bar{\cal W}),\cr
\varphi =i\log\sqrt{\frac{\bar{\cal W}}{{\cal W}}}= i\log\sqrt{\frac{\bar{\cal V}}{{\cal V}}}.\label{4-6}
\end{eqnarray}

We can write   
\begin{equation}
dz^2+d\mathpzc{y}^2+dx^2+dy^2=d{\cal V}d\bar{\cal V}+d{\cal W}d\bar{\cal W}.\label{4-7} 
\end{equation}

We now have $|{\cal V}|^2 +|{\cal W}|^2=x^2+y^2+z^2+\mathpzc{y}^2=r^2+R^2$.
So we identified $\mathds{C}^1\times\mathds{C}^1$ with $\mathds{R}^4$ and so contains $S^3$, given by $| {\cal V}| ^2 +|{\cal W}|^2=const.$ 
Every line through the origin, represented by $({\cal V}, {\cal W})$ intersects the sphere $S^3$, for example $(\lambda{\cal V}, {\lambda\cal W})$ with $\lambda=\frac{1}{\sqrt{\mid {\cal V}\mid ^2 +\mid{\cal W}\mid^2}}$. Thus the homogeneous coordinates can be restricted to $\mid {\cal V}\mid ^2 +\mid{\cal W}\mid^2=1.$ 

The point  $({\cal V}, {\cal W})\in S^3\subset\mathds{C}^1\times\mathds{C}^1$ with $| {\cal V}| ^2 +|{\cal W}|^2=1$, becomes by the complexification,  a point of $S^2$, so with the single complex coordinate ${\cal Z}=\frac{{\cal V}}{{\cal W}}$. We have now a map $H:S^3\rightarrow S^2$, which is continuous. 
One calls this a Hopf map.
For each point of $S^2$, the coordinate $({\cal V}, {\cal W})$ is non unique, because it can be replaced by $(\lambda{\cal V}, \lambda{\cal W})$, such that $|\lambda|^2=1, \lambda\in S^1$.
So we will now write $\mathds{C}_1$ for $S^2/\{\infty\}$ and $\mathds{C}_2$ for $S^2/\{0\}$ and admitting coordinates ${\cal Z}$ and ${\cal Z}'=\frac{1}{{\cal Z}}$ respectively. Let $G$ be the group of self-homeomorphisms of the product space $S^2\times S^2$, generated by interchanging the two coordinates of any point and by the antipodal map on either factor. 
$G$ is then isomorphic to the  dihedral group. It contains 3 subgroups, for example $K=\{I,(x,y)\rightarrow (-x,y), (x,y)\rightarrow (x,-y), (x,y)\rightarrow (-x,-y)\}$. It acts freely on $S^2\times S^2$. Then $(S^2\times S^2)/K=\mathds{R}P^2\times\mathds{RP}^2$.
The most interesting feature is the fact that the 2-fold symmetric product of $\mathds{R}P^2$,  $SP^2 (\mathds{R}P^2)=\mathds{R}P^4$.
\subsection{Hopf fibrations of the Klein surface\label{sec:4.3}}
Consider the quadratic Hopf map $ f_H: S^3\rightarrow S^2$ given by\cite{Steenrod1951}. 
\begin{equation}
f_H({\bf z,w})=(|{\bf z}|^2-|{\bf w}|^2,2{\bf z}\bar {\bf w}).\label{4-8}
\end{equation}
This map can be compared with the homomorphism $SU(2)\rightarrow SO(3)$ (or with O(3), if we allow antipodal mappings). Remember that we have also $SO(3)\cong{\cal M}_0(\mathds{C}_\infty)=SU(2)/\{\pm I\}$.
$SO(3)$ acts transitively on $S^2$ and can be represented by a rotation  $R_{\theta ,x}$ over $\theta$ at an  axis  $\mathds{R}x$. So $SO(3)_x$ acts on the tangent space $T_x(S^2)$ via rotations.
Further, $\mathds{R}P^2$ can minimally embedded in $S^4$ (or $\mathds{C}^2$). This is sometimes called a "Veronese surface".
The components of the Veronese map $S^2\rightarrow S^4$ of degree 2 are homogeneous polynomials on $\mathds{R}^3$. For the Cartesian coordinates $(x_1,x_2,x_3)\in S^2\subset \mathds{R}^3$ one then can write
\begin{equation}
Ver_2({\bf x})=C\Bigl(x_1{\bf zw}-x_2\frac{{\bf z}^2-{\bf w}^2}{2}+ix_3\frac{{\bf z}^2+{\bf w}^2}{2}\Bigr)^2.\label{4-9}
\end{equation}
This map $Ver_2$, factors through the antipodal map, resulting in a minimal immersion of $\mathds{R}P^2$ into $S^4$. $Ver_2$ is conformal with conformal factor $C$.
The veronese map in components is
\begin{equation}
Ver_2({\bf x})=C\Bigl(\sqrt{3}(x_2^2-x_3^2),-2\sqrt{3}x_2x_3,-2\sqrt{3}x_1x_2,2\sqrt{3}x_1x_3,2x_1^2-x_2^2-x_3^2\Bigr).\label{4-10}
\end{equation}
The harmonicity can be verified by calculating $\frac{\partial^2}{\partial x_i^2} (Ver_2({\bf x}))$\cite{Toth2002}.

Next, we make the connection with the Klein bottle.
The Klein bottle fibration is locally trivial, but not globally.
Summarized, the Hopf fibration of the 3-sphere in $\mathds{R}^4$ can be depicted as
\begin{eqnarray}
\begin{tikzcd}
& ({\cal V}, {\cal W})\:\:\:\:\: \in \hspace{-1.5cm}  \arrow[d, "H"]
& \mathds{C}^2/\{0\} \arrow[d, "\mathds{C}P^1"]
& S^3 \arrow[l]\arrow[d] \\
& \:\:\:\:\:{\cal Z}\equiv\frac{{\cal V}}{{\cal W}} \:\:\:\:\:\: \:\: \in\hspace{-1cm}
&\:\:\:\: \mathds{C}\cup\{\infty\}\:\:\:\:  =\hspace{-1cm}
& S^2
\end{tikzcd}\label{4-11}
\end{eqnarray}
In our model, without the 5D contribution (see eq.(\ref{2-6}), the polynomial for $D_1=0$ is indeed of order 2 on $S^2$. and becomes via the Veronese map a polynomial of order 5.

We will continue with the extended theory of fibrations. The interested reader can consult for example Steenrod \cite{Steenrod1951} or an interesting study by Urbantke \cite{ur2003}.

\section{Minimal  non-orientable  surfaces, half-spaces and purification of Hawking radiation\label{sec:5}}
\subsection{On the minimal Klein surface\label{sec:5.1}}
A question still remains: is our solution  related to the complete minimal non-orientable Klein surface  in $\mathds{R}^3$ with finite total curvature,  whose generalized Gauss map omits one point in $\mathds{R}P^2$ (two points on the sphere).
The existence of lowest total curvature can be of interest concerning stability and  conformal structure of the connected Riemann surface, say $(M,ds^2)$.
One can have harmonic, holomorphic and meromorphic (in our case) functions on $S^2$.
Using the complex structure, one can apply the Weierstrass representation\cite{martin2001} in order to construct  a conformal minimal immersion $X:M\rightarrow \mathds{R}^3$ (with $Y:M\rightarrow S^2$ its Gauss map).
An immersion of a non-orientable minimal surface is minimal if the mean curvature of $X$ on any orientable piece of $M$ is zero.
\begin{figure}[h]
\centering
\resizebox{0.6\textwidth}{!}
{\includegraphics{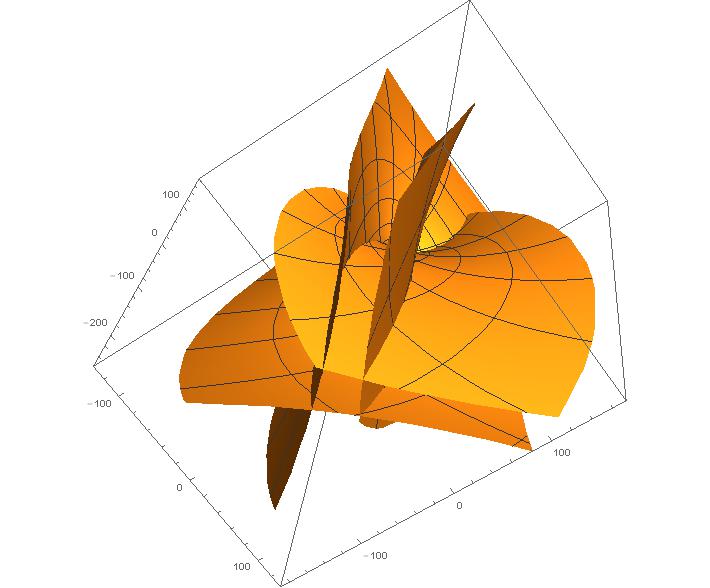}}
\caption{Plot of the minimal surface of  $F$ (eq.(\ref{3-5}))}
\label{fig:10}       
\end{figure}
The only complete non-orientable minimal surface of the one-ended Klein bottle with total curvature $-8\pi$ is plotted\cite{lopez1993}. It has the well known self-intersection. However, it can be embedded in $\mathds{R}^4$.
Complete non-orientable minimal immersion delivers restrictions on the topology of the original manifold, such as our  horizon model in the antipodal setting and so on the embedding in $\mathds{R}^4$ (coordinates $(z,\mathpzc{y}, r,\varphi)$. Note that in our 5D manifold,  we still have the time  coordinate, which can be absorbed in the Eddington-Finkelstein coordinate $U$ or $V$. 
Further, during the evaporation process of the black hole, it is reasonable to assume that the surface will become minimal.
Minimal non-orientable surfaces, such as  Meeks M\"obius strip, can possess instabilities under  parametric variation  of its boundary\cite{pesci2015}. 

There is a strong theorem for a  non-orientable minimal surface\cite{hoffman1990}.
Suppose that $M$ is a proper, possibly branched connected minimal submanifold of $\mathds {R}^n$ whose boundary $\partial M$ (which may be empty), is  a compact set. Let $H(M)$ denote the {\it convex hull} of $S$. Then $H(M)$ is a  half-space, for example the  half-catenoid or a hyperplane. For $n=3$, $\partial M$ has nonempty intersection with each boundary component of $H(M)$. This is needed for our two hemispheres.
Further, the only complete non-orientable minimal surface with total curvature $-8\pi$, is the one-ended Klein bottle. It is conjectured, that when the total curvature is $-6\pi$, the moduli space is discrete. For example the projective space $\mathds{R}P^n$ or $\mathds{C}P^n$, which parametrizes  lines in $\mathds{R}^{n+1}$ and $\mathds{C}^{n+1}$.
Entering the other "half-"hemisphere (in the Penrose diagram, region II), one applies then  the antipodal mapping.

Let $X: M\rightarrow \mathds{R}^3$ be a complete minimal immersion of a  non-orientable Riemannian surface  in $\mathds{R}^3$ with finite total curvature. Let $J_0: M_0\rightarrow M$ be the conformal oriented two sheeted cover of $M$ and $I: M_0\rightarrow M_0$ the anti holomorphic order-two deck transformation for this cover. In our case   the antipodal map  $S^{n-1}\times\mathds{R}\rightarrow S^{n-1}\times\mathds{R}, (U,V,z,\mathpzc{y},\varphi )\rightarrow (-U,-V,-z,-\mathpzc{y},\varphi+\pi )$.
Then $X_0=X\circ J_0$ is a minimal immersion of an orientable surface.
We denote with $(M,I,g,\eta)$ the Weierstrass representation of a orientable minimal immersion of $X_0$. 
Here $(g,\eta)$ represents the Weierstrass data (or M($\Phi$), with $\Phi_i$ the well known holomorphic Weierstrass 1-forms, which can be expressed in $g,\eta$) of $X_0$. 

The Weierstrass representation for $X$ can then be formulated by the anti-holomorphic involution (or antipodal map) $\mathpzc{f}$ in the Riemann sphere $\mathds{C}_\infty$: $\mathpzc{f}({\bf z})=-\frac{1}{\bar {\bf z}}\in \mathds{C}$. One then writes $J: \mathds{\bar C}\rightarrow\mathds{R}P^2= \mathds{\bar C}/\{1,\mathpzc{f}\}$ (the two-sheeted cover of the projected plane).
Further, one has the properties $\mathpzc{g}\circ I=\mathpzc{f}\circ  \mathpzc{g}$,   $I^*(\Phi_i)=\bar\Phi_i$ and $\overline{I^*(\eta)}=-\eta g^2$ ($g$ is the Gauss map of $M_0$)
One defines then the Gauss map of the non-orientable immersion $X$ as the conformal map $G: M\rightarrow \mathds{R}P^2$ with $G\circ J_0= J\circ g$.
If one considers the torus
\begin{equation}
\bar M^{tor}_r=\{(z,u)\in \mathds{C}_\infty\times\mathds{C}_\infty;\quad  z^2=\frac{u(u-r)}{ur+1}\}\quad r\in\mathds{R},\label{5-1}
\end{equation}
with $I$  the anti holomorphic involution without fixed points: $I(z,u)=(\frac{1}{\bar z},-\frac{1}{\bar u})$ and $M _r=\bar M^{tor}_r/\{0,\infty\}$ a Riemann surface of genus 1, then $M_r'=M_r/<I>$ represents the non-orientable conformal surfaces\cite{lopez1993}.
Note that  the Weierstrass representation is comparable with the complex map on the Riemann sphere.
The function $g$ is the meromorphic function $F$ of eq.(\ref{3-4}). There is no objection to extend the representation to $\mathds{R}^4$ were the map becomes a embedding in stead of a immersion, because the meromorphic function is independent of $\mathpzc{y}$.

Another proven method to describe minimal surfaces, can be delivered by the introduction of twistors\cite{penrose1984}. In figure 9 we plotted the minimal surface of our quintic.
\subsection{Convex hull and the quintic\label{sec:5.2}}
Let us return to the principal form of our quintic, eq.(\ref{3-12}).
Suppose that we can write  $F(z)=(z-z_1).....(z-z_5)$. Then
\begin{equation}
\frac{F'}{F}=\frac{1}{(z-z_1)}+...+ \frac{1}{(z-z_5)}.\label{5-2}
\end{equation}
One proves\cite{pras2001} with the help of eq.(\ref{5-2}), that the roots of $F'$ belong to the convex hull of the roots of $F$. In our case the roots of $F'\sim r(r-a)^3$ are $(0,a)$. See eq.(\ref{2-5}).
It is conjectured that in our model all the singularities will lie inside the hull. 
Further, we have in our whole spacetime $r\geq a$. By the antipodal identification, two points at distances $<a$ are identified. Only for $r=0$ we would have a problem, however it occurs in the infinite  far future for the outside observer.
Conclusive, by the conformal description,  $^{(4)}\bar g_{\mu\nu}$ is conformally flat without any essential singularities.
\subsection{Connection with purification\label{6.3}}
Suppose we have the convex hull of a set of points (particles), which can be dynamical. This is the case for the in-going and out-going  Hawking particles near the horizon. The set of points coincide  then with the total number of particles emitted during the evaporation of the black hole.
As already remarked in the foregoing sections and the introduction, the in-going particles emerge as out-going particles at the antipode (region II in the Penrose diagram, which are two opposite sides of the same black hole). Cusp singularities are then avoided\footnote{Remember that also the variables $u^\pm, p^\pm$ all switch signs from I to II.}. In fact, regions I and II represent together the state the black hole is in. 
Our 5D spacetime will then be cut out by a $(\mathds{C}^1\times\mathds{C}^1)$ 4-sphere and the antipodal identification  $(U,V,z,\varphi,\mathpzc{y})\rightarrow (-U,-V,-z,\varphi+\pi,-\mathpzc{y})$ is applied.
But the major change, as described by 't Hooft\cite{thooft2018}, is that the out-going Hawking radiation will remain pure states for the outside observer in stead of a thermodynamically mixed state. 
Pure quantum states of the local observer map onto pure quantum states of the black hole as seen by a distant observer.
The former "hidden sector" is actually the other side of the black hole. So we are dealing with only pure states.
This is possible, because the model is invariant under $\Omega$ (eq.(\ref{2-3})). So different observers use different $\Omega$, because we could write the Ricci scalar 
\begin{equation}
\bar R=\frac{18}{\bar\omega^2N^2}\Bigl(\dot{\bar\omega}^2-N^4\bar\omega'^2\Bigr)-\frac{2}{3}\kappa^2\Lambda \bar\omega^2.\label{5-3}
\end{equation}
The in-going observer will choose $\Omega$ such that $\bar R=0$.
Our conformal invariant 5D warped spacetime, invariant under $\Omega$, has a conformally flat 4D effective spacetime, where the difference in experience of the vacuum state for the in- and out-going observer is expressed in the dilaton field. 
It was already Schr\"odinger, who presented in 1936 a study  about "the realization of a mixed state of a quantum state as an ensemble of pure quantum states and the relation  between the purification of the density matrix"\cite{schrod1936} and nowadays known as the  Schr\"odinger-HJW theorem\cite{kirk2006}. It states that any mixed state can be written as a convex combination of pure states. The mixed states are in the two half hemispheres  (region I and II in the Penrose diagram) of one black hole.  
In this representation, orthogonal states correspond to antipodal points of the complex Riemann sphere (sometimes called, in context of qubit states, the Bloch sphere). The original  elliptic interpretation of Schr\"odinger fits very well now\cite{schrod1957}.
The points $P(X)$ and $P(\bar X)$ in figure 2 represent the same world-point. They are joined by a  spacelike geodesic and  the lightcones at $P(X)$ and $P(\bar X)$ have no points in common. Any half  hemisphere represents the whole world when it contains no antipodal pairs.
The distinction between past and future is lost as well as the  fore-cone and after-cone. This  is just the antipodal identification $(U,V,z,\mathpzc{y},\varphi )\rightarrow (-U,-V,-z,-\mathpzc{y},\varphi+\pi )$. Time will also reverse from $I$ to $II$\footnote{The original considerations of Schr\"odinger is worth reading!}.
However, an observer outside will of course notice a thermodynamical system. By measuring entropy, he observes an arrow of time.  
So points on our complexified $\mathds{R}^4$ (see eq.(\ref{4-6})) become, by the Hopf fibration, eq.(\ref{4-11}),  points in a two-sphere with antipodal points corresponding to a pair of mutual orthogonal state vectors. 
\section{Summary\label{sec:sum}}
The exact time-dependent solution of a black hole on a five-dimensional warped spacetime in conformal dilaton gravity is throughout investigated. The level of badness of the singular points are determined by a quintic equation. The investigation by means of a complex analysis, shows that there are no essential singularities.
A modification of the topology of the $(4+1)$-dimensional warped spacetime is necessary , i.e., the identification of antipodal points on the horizon. This is mandatory in order to guaranty unitarity and quantum pureness. This alteration, which was proposed by 't Hooft in the 4D case, will also solve the firewall problem and information paradox and maintain CPT inversion.
The antipodal identification removes the inside of the black hole, only notable for the outside observer. So one will not enter "region II" of the Penrose diagram. This region refers to the  opposite of the hemisphere of the same black hole.
It is conjectured that the quantum mechanical issues concerning the particles transiting information across the horizon, is still linked with the gravitational back reaction.
In the original 4D model, the map between the entire asymptotic domain and ordinary spacetime   must be one-to-one (in order to preserve the metric) and was accomplished by identifying the antipodal map with  the element $-\mathds{I}\in \mathds{Z}_2$, subgroup of $O(3)$.  
Now we have in our 5D warped model $\mathds{Z}_2$ symmetry with  respect to the extra dimension  $\mathpzc{y}$. So we can maintain the antipodal mapping of the horizon onto itself without fixed points. The only difference is that we must consider the Klein surface in stead of the M\"obius strip.
We  used the  $\mathds{Z}_2$ symmetry in the $\mathpzc{y}$-coordinate and made a connection between the polynomial solution and the complex description of the $\mathds{R}^4$, needed for the embedding of the Klein surface.
In order to arrive at the cut-out of the interior of the 3-sphere  $S^3$ of the black hole, we must apply the Hopf-map of the $S^4$, in complex coordinates $\mathds{C}^1\times\mathds{C}^1$, via the complex projective planes.
It is found that the structure of the singularities of the exact black hole solution, presented here by the quintic polynomial,
possesses a deep-seated connection with the conformal d'Alembert and the icosahedron symmetry group.
It is conjectured that a warped 5D description is necessary. 
A connection was made with the construction of minimal surfaces in $\mathds{R}^4$ of the Klein bottle. They play an important role during the evaporation process of the black hole.
Finally, the purification of the outgoing particle is described in our model.
\section*{\centerline{Appendix A.}\\ \centerline{The icosahedron group}}
\setcounter{equation}{0}
\renewcommand{\theequation}{A-\arabic{equation}}
Let us  inscribe the icosahedron in $S^2$, with the north and south pole vertices. The icosahedran is made up of a north and south pentagonal pyramid separated by a pentagonal antiprism. See figure 14. 
\begin{figure}[h]
\centering
\resizebox{0.9\textwidth}{!}
{\includegraphics{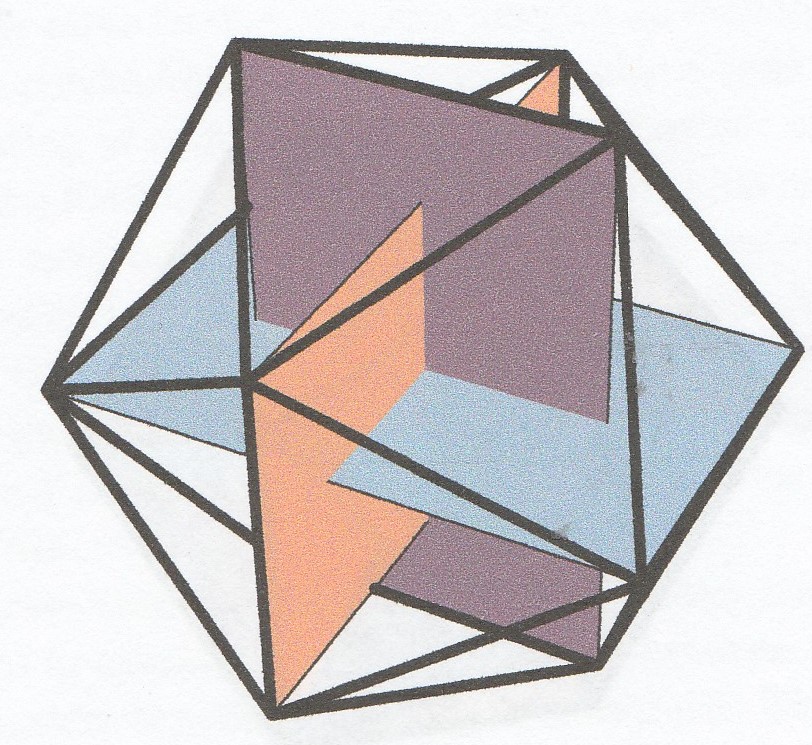},\includegraphics{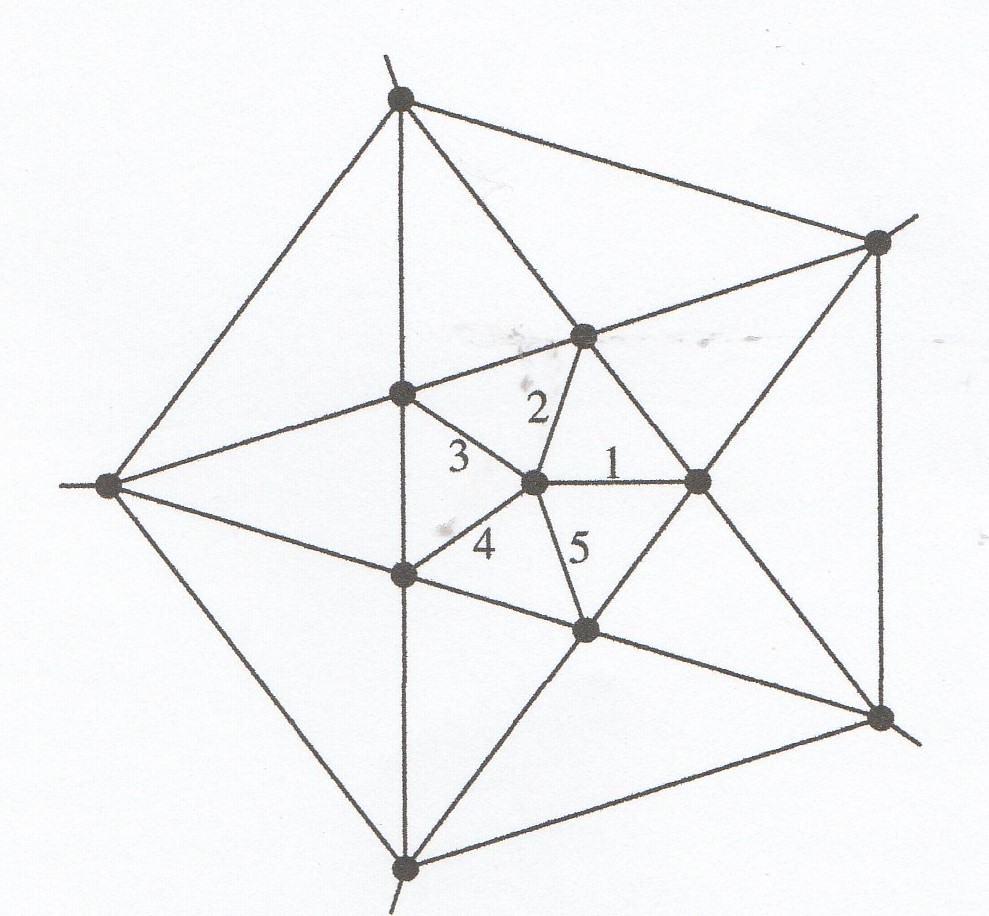}}
\caption{  Left: The icosahedron.  Right: stereographic projected vertices, where two of them are at the poles. The generators are a one-fifth rotation about the north pole. These are the most interesting ones of the sixty of the ${\cal A}_5$ group. The 5 depicted vectors are just the 5 roots 
of the black hole solution in the complex plane. See also figure 3.}
\label{fig:14}       
\end{figure}

Let $G$ be a finite group of isometries of $\mathds{R}^3$. Each element $ \mathpzc{g}$ of $G$ is a rotation. A rotation axes through $O$ intersects $S^2$ in two antipodal points. We denote all the sets of antipodal points as $\mathpzc{P}$. 
The isometry group $G_{\mathpzc{g}}=\{{\mathpzc{g}}\in G\mid {\mathpzc{g}}(x)=x\}$ for each pair of antipodal points, is cyclic.  The rotations are a multiple of the smallest angle of rotation. So $G$ has a degree, say, of order $d$.
The icosahedron is of order 5 and has 3 orbits in $\mathpzc{P}$ with total 20, 12 and 30  elements. Each orbit is invariant under the antipodal map. The direct group of isometries of $\mathds{R}^3$ for the icosahedron is the group ${\cal A}_5$ (for a detailed treatment, see Toth \cite{Toth2002}).
The group $G$ is linear, because the only fixed point is the origin $O$. The group of all linear isometries of $\mathds{ R}^3$ is the orthogonal group $O(3)$ of orthogonal $(3\times 3)$ matrices, with determinant $\pm 1$. For the positive determinant we have the $SO(3)$. We need the $O(3)$ for the reflections through $O$. Further, there is a bijective correspondence between $SO(3)$ and $\mathds{R}^3$.

Now we will consider again  the linear fractional M\"obius transformations as the linear rotations ${\cal R}_{\varphi ,x}$ around an axis through two antipodal points $(x,x')$ over angle $\varphi$, element of $O(3)$ ($x=(x_1,x_2,x_3)$). Then, by Cayley's theorem, ${\cal R}_{\varphi ,x}$ conjugated with the stereographic projection $h$ is the linear fractional transformation
\begin{equation}
(h\circ {\cal R}\circ h^{-1})(\zeta)=\frac{z\zeta-\bar w}{w\zeta +\bar z},\qquad \zeta \in \mathds{C}_\infty,\label{A-1}
\end{equation}
where 
\begin{equation}
z=\cos\Bigl(\frac{\varphi}{2}\Bigr)+i\sin\Bigl(\frac{\varphi}{2}\Bigr) x_3,\quad w=\sin\Bigl(\frac{\varphi}{2}\Bigr)(x_2+ix_1).\label{A-2}
\end{equation}

The two fixed points  $\frac{x_3\pm 1}{x_1-ix_2}$ are just the antipodal points $(x,x')$.
Further, we have the isomorphisms  $O(3)\cong{\cal M}_0(\mathds{C}_\infty) =SU(2)/{\pm I}$.
Now we want to find  finite M\"obius groups $G\subset{\cal M}(\mathds{C}_\infty)$ and to construct rational functions invariant under the M\"obius groups.

Subgroups of G can be linearized by lifting  it to $SU(2)$ along $2:1$ projection: $SU(2)\rightarrow {\cal M}_0(\mathds{C}_\infty)$. This in the binary cover group  $G^*$ of $G$. The reason is, that subgroups of $SU(2)$ can easily be found.
The goal is to solve the invariance issue for homogeneous polynomials for $G^*$ , $\mathds{C}^2\rightarrow \mathds{C}$ and to prove the relation between the quintic and the icosahedron symmetries.
Quotients of some of these invariant polynomials define holomorphic self-maps in the complex  projected plane  $\mathds{C}P^1=\mathds{C}_\infty$. These quotients restrict to rational functions on $\mathds{C}$ and obey the invariance group $G$.
One can compare the singularities of these rational functions with the roots of the invariant form of the icosahedron group  ${\cal A}$.

Let us consider the  cyclic group $C_d$ of order $d$,  isometric to $\mathds{R}^3$ and representing rotations of $\mathds{C}_\infty$:
\begin{equation}
\zeta\rightarrow e^{2\pi i\frac{l}{d}}\zeta,\qquad l=0, ... ,d-1.\label{A-3}
\end{equation}
It is the group of rotations with rotation axis through the north and south pole.
If we adjoin to $C_d$ then inversions $\zeta\rightarrow\frac{1}{\zeta}$, we obtain the  dihedral M\"obius group $D_d$
\begin{equation}
\zeta\rightarrow \Bigl(e^{2\pi i\frac{l}{d}}\zeta, \frac{1}{\zeta}e^{-2\pi i\frac{l}{d}}\Bigr),\qquad l=0, ... ,d-1.\label{A-4}
\end{equation}
Note that they are obtained by taking $z=e^{\pi i\frac{l}{d}}, w=0$ and $w=ie^{\pi i\frac{l}{d}}, z=0$ respectively in eq.(\ref{A-2}).
For the icosahedron M\"obius group $I$, by suitable orientation of the axes (figure 14), we obtain for the linear fractional transformations
\begin{equation}
\zeta\rightarrow \omega^l\zeta,\qquad l=0, ...,4,\label{A-5}
\end{equation}
with $\omega =e^{2\pi i\frac{1}{5}}$. Then it straightforward to find all the 60 elements of $I$ expressed in $\omega$ \cite{Toth2002}. The most interesting elements are the vertices of the icosahedron, 5-fold stereographically projected on  $\mathds{C}_\infty$ are
\begin{equation}
0,\quad \infty,\quad \omega^l(\omega+\omega^4),\quad \omega^l(\omega^2+\omega^3).\label{A-6}
\end{equation}
If one takes the inverse image, $SU(2)\rightarrow {\cal M}_0(\mathds{C}_\infty)$, we obtain the binary group $G^*$. Taking the inverse images from the finite M\"obius group inn the case of the icosahedron, we obtain the binary icosahedron group $I^*$:
\begin{eqnarray}
I^*=\Bigl\{\pm\omega^l,\pm i\omega^l, \pm\frac{1}{\sqrt{5}}\Bigl(-\omega^{3l}(\omega-\omega^4)+i\omega^{2l}(\omega^2-\omega^3)\Bigr)\omega^{3k},\cr
\pm\frac{1}{\sqrt{5}}\Bigl(\omega^{3l}(\omega^2-\omega^3)+i\omega^{2l}(\omega-\omega^4)\Bigr)\omega^{3k}\Bigr),\qquad k, l=0,...,4. \Bigr\}\label{A-7}
\end{eqnarray} 

Consider now the $S^3$,  parametrized by a  quaternion  $z+iw, z,w\in \mathds{C}, |z| ^2+ |w|^2=1$. 
If we work for the moment in spherical polar coordinates, we can parametrize $z=\cos\tau e^{i\theta}$ and $w=\sin\tau e^{i\varphi}, 0<\tau<\pi/2$. One defines for fixed $\tau$:
\begin{equation}
T_\tau =\{(p,q)\in S^3\subset \mathds{C}^2,\quad |z|^2-|w|^2=\cos 2\tau\}.\label{A-8}
\end{equation}
For $\tau =0, \pi/2$ we have two circles cut out from $S^3$.
We can choose for the coordinates in these sub spaces $(x_1,x_2)$ and $(x_3,x_4)$ in $\mathds{R}^4=\mathds{C}^2$. In our situation, we could have  $(r, z), ( \varphi, \mathpzc{y})$.
$T_\tau$ represents a torus, or a Klein bottle by suitable identification of squares $[0,2\pi]\times[0,2\pi]$. We know that a Klein bottle can be embedded in $\mathds{R}^4$.
The cyclic and binary describes  groups above, fits very well in this Clifford decomposition of $S^3=SU(2)$. All the elements of $I^*$ can be projected on the two Klein surfaces by suitable choices of $\cos(2\tau)$.
So every finite subgroup of $S^3$ is either cyclic or conjugate to the binary subgroups, in our case the $I^*$. 

Before we arrive at the  quintic polynomial on $\mathds{R}^4$ as solution of the conformal Laplacian in 4 space components, we first define  invariant forms  of the binary polyhedral groups, i.e., in our case for the icosahedron, of degree 5.

We found that $SL(2,\mathds{C})$ acts on $\mathds{C}^2$ by matrix multiplication.
For $g\in SL(2,\mathds{C})$  and $(z,w)\in \mathds{C}^2$, 
\begin{equation}
(z,w)\rightarrow g.(z,w)=\frac{az+bw}{cz+dw}=\frac{a\zeta+b}{c\zeta+d},\quad \zeta=\frac{z}{w}.\label{A-9}
\end{equation}
The action of $g$ extends to the projective space $\mathds{C}P^1$ by setting $[z:w]\rightarrow g[z:w]$
by using homogeneous coordinates on $\mathds{C}P^1$.
On his turn, $\mathds{C}P^1$ can be identified with $\mathds{C}_\infty$ by $[z:w]\rightarrow \zeta=\frac{z}{w}, [z:w]\in \mathds{C}P^1$.
So summarized, the action of $G$ on  $\mathds{C}_\infty$ by linear fractional transformations, corresponds to actions of $G^*\subset SL(2,\mathds{C})$ on $\mathds{C}^2$ by matrix multiplication.

Further, the automorphisms of the sphere are fractional linear transformations, represented as projective matrix classes, i.e., rotation groups such as the ${\cal A}_5$ for the icosahedron group.

\section*{\centerline{Appendix B. }\\ \centerline{The Einstein equations written out in components}}
\setcounter{equation}{0}
\renewcommand{\theequation}{B-\arabic{equation}}
We used the algebraic Maple program GRtensorIII in order to obtain the field equations. We checked the equations with the help of the Mathematica program OGRe.
The 5D Einstein equations equations are

\begin{eqnarray*}
\hspace{-2cm}\underline{{\bf Eins}_t^t}: \Bigl(\frac{3}{16}\Bigr)^{\frac{2}{3}}\kappa^{\frac{4}{3}}\Lambda\omega^{\frac{10}{3}}+\frac{N\omega}{r}N'(\omega+2r\omega')+ +\frac{1}{4}r^2\omega^2{N^{\varphi'}}^2 
+\frac{3}{2}\omega^2rN^\varphi {N^\varphi}'+\frac{1}{2}\omega^2r^2N^\varphi {N^\varphi}''\cr
\hspace{-1.5cm}+\omega r^2\omega' N^\varphi {N^\varphi}'+2\frac{\omega}{r}N^2\omega'+2\omega N^2\omega''+2\frac{\omega}{N^3}\dot N\dot\omega-\frac{8}{3N^2}\dot\omega^2-\frac{2}{3}N^2\omega'^2\cr
\hspace{-2cm}\underline{{\bf Eins}_t^r}: \omega N(2N'\dot\omega-\dot N(\frac{\omega}{r}+2\omega'))-\frac{1}{2}N^\varphi  r^2\omega (2N_\varphi'\dot\omega+\omega \dot N^{\varphi'})
+\frac{2}{3}N^2(5\omega'\dot\omega-3\omega{\dot\omega}')\cr
\hspace{-2cm}\underline{{\bf Eins}_\varphi^t}: \frac{1}{2}r\omega(2r{N^\varphi}'\omega'+r\omega{N^\varphi}''+3\omega{N^\varphi}')\cr
\hspace{-2cm}\underline{{\bf Eins}_r^r}: \Bigl(\frac{3}{16}\Bigr)^{\frac{2}{3}}\kappa^{\frac{4}{3}}\Lambda\omega^{\frac{10}{3}}+\frac{1}{4}r^2\omega^2{N^\varphi}^2+\frac{N\omega N'(\omega+2r\omega')}{r}
+\frac{2N^2\omega'(3\omega+4r\omega')}{3r}\cr+\frac{2\omega\dot N\dot\omega}{N^3}+\frac{2(\dot\omega^2-3\omega\ddot\omega)}{3N^2}\cr
\hspace{-2cm}\underline{{\bf Eins}_\varphi^r}: -\frac{1}{2}r^2\omega(2\dot\omega {N^\varphi}'+\omega \dot {N^\varphi}')\cr
\hspace{-2cm}\underline{{\bf Eins}_z^z={\bf Eins}_y^y}:\Bigl(\frac{3}{16}\Bigr)^{\frac{2}{3}}\kappa^{\frac{4}{3}}\Lambda\omega^{\frac{10}{3}}+\omega^2\Bigl(N'^2 +N(\frac{2N'}{r} +N'') -\frac{3\dot N^2}{N^4}+\frac{\ddot N}{N^3}
-\frac{1}{4}r^2{N^\varphi}'^2 \Bigr)\hspace{2.8cm}\cr+\frac{2(\dot\omega^2-N^4\omega'^2)}{3N^2}+\frac{2\omega\Bigl(2rN^4N'\omega'+N^5(\omega'+r\omega'')+2r\dot N\dot\omega-rN\ddot\omega\Bigr)}{rN^3}\cr
\hspace{-2cm}\underline{{\bf Eins}_\varphi^\varphi}: \Bigl(\frac{3}{16}\Bigr)^{\frac{2}{3}}\kappa^{\frac{4}{3}}\Lambda\omega^{\frac{10}{3}}+\omega^2\Bigl( N'^2 +\frac{N^5N''-3\dot N^2+N\ddot N}{N^4}
-\frac{r}{4}(3r{N^\varphi}'^2+2N^\varphi(3{N^\varphi}'+r{N^\varphi}''))\Bigr)\hspace{1.4cm}\cr+\frac{2(\dot\omega^2-N^4\omega'^2)}{3N^2}
+\omega\Bigl( 4NN'\omega'-N^\varphi r^2{N^\varphi}'\omega'+2N^2\omega''+\frac{4\dot N\dot\omega}{N^3} -\frac{2\ddot\omega}{N^2}\Bigr)
\end{eqnarray*}

The effective 4D Einstein equations are  collected below:

\begin{eqnarray*}
\hspace{-2cm}\underline{{\bf Eins}_t^t}: \frac{1}{6}\kappa^2\Lambda \omega^2+\frac{N}{r\omega}N'(\omega+2r\omega') +\frac{1}{8}r^2{N^{\varphi'}}^2+rN^\varphi {N^\varphi}'
+\frac{1}{3}r^2N^\varphi {N^\varphi}''+\frac{r^2}{\omega}\omega' N^\varphi {N^\varphi}'\cr\hspace{-1.5cm}+\frac{2}{r\omega}N^2\omega'+\frac{2}{\omega}N^2\omega''+\frac{2}{N^3\omega}\dot N\dot\omega
-\frac{3}{N^2\omega^2}\dot\omega^2-\frac{1}{\omega^2}N^2\omega'^2+\frac{1}{6}NN''+\frac{1}{6}N'^2
+\frac{1}{6}\frac{\ddot N}{N^3}-\frac{1}{2}\frac{\dot N^2}{N^4}\cr
\hspace{-2cm}\underline{{\bf Eins}_t^r}:\frac{N}{\omega}(2N'\dot\omega-\dot N(\frac{2\omega}{3r}+2\omega'))-\frac{1}{3\omega}N^\varphi  r^2(3N_\varphi'\dot\omega+\omega \dot N^{\varphi'})
+\frac{2}{\omega^2}N^2(2\omega'\dot\omega-\omega{\dot\omega}')\cr
\hspace{-2cm}\underline{{\bf Eins}_\varphi^t}: \frac{r}{3\omega}(3r{N^\varphi}'\omega'+r\omega{N^\varphi}''+3\omega{N^\varphi}')\cr
\hspace{-2cm}\underline{{\bf Eins}_r^r}: \frac{1}{6}\kappa^2\Lambda\omega^2+\frac{1}{8}r^2{N^\varphi}^2+\frac{N N'(\omega+2r\omega')}{r\omega}+\frac{N^2\omega'(2\omega+3r\omega')}{r\omega^2}
+\frac{2\dot N\dot\omega}{\omega N^3}\hspace{2.5cm}\cr+\frac{(\dot\omega^2-2\omega\ddot\omega)}{N^2\omega^2}+\frac{1}{6}\Bigl(NN''+N'^2+\frac{\ddot N}{N^3}-3\frac{\dot N^2}{N^4}\Bigr)\cr
\hspace{-2cm}\underline{{\bf Eins}_\varphi^r}:-\frac{r^2}{3\omega}(3\dot\omega {N^\varphi}'+\omega \dot {N^\varphi}')\cr
\hspace{-2cm}\underline{{\bf Eins}_z^z}:\frac{1}{6} \kappa^2\Lambda\omega^2+\frac{5}{6}\Bigl(N'^2 +N(\frac{2N'}{r} +N'')\Bigr)-\frac{5\dot N^2}{2N^4}+\frac{5\ddot N}{6N^3}\cr
-\frac{5}{24}r^2{N^\varphi}'^2 +\frac{(\dot\omega^2-N^4\omega'^2)}{\omega^2N^2}+\frac{2\Bigl(2rN^4N'\omega'+N^5(\omega'+r\omega'')+2r\dot N\dot\omega-rN\ddot\omega\Bigr)}{r\omega N^3}\cr
\hspace{-2cm}\underline{{\bf Eins}_\varphi^\varphi}: \frac{1}{6} \kappa^2\Lambda\omega^2+\frac{5}{6}\Bigl( N'^2 +\frac{N^5N''-3\dot N^2+N\ddot N}{N^4}\Bigr)+\frac{1}{3r}NN'+\frac{(\dot\omega^2-N^4\omega'^2)}{\omega^2 N^2}\cr
\hspace{-1.5cm}-\frac{r}{3}(\frac{13}{8}r{N^\varphi}'^2+N^\varphi(3{N^\varphi}'+r{N^\varphi}''))
+\frac{1}{\omega}\Bigl( 4NN'\omega'-N^\varphi r^2{N^\varphi}'\omega'+2N^2\omega''+\frac{4\dot N\dot\omega}{N^3} -\frac{2\ddot\omega}{N^2}  \Bigr)
\end{eqnarray*}
The differences are caused by the contribution from the projected Weyl tensor orthogonal to $n^\mu$.
By combining the several components, one easily obtains the dilaton equation  for the 5D case as well as for the 4D case.


\section{References}\label{refs}
\begin{thebibliography}{99}
\bibitem{Hawking1975}
S Hawking: {\it Particle creation by black holes},  Comm. Math. Phys. {\bf 43} (1975) 199.
\bibitem{suss1993}
L Susskind, L Thorlacius and J Uglum. {\it The stretched horizon and black hole complementarity}, Phys. Rev. D {\bf 48} (1993) 1743.
\bibitem{page1993}
D N Page. {\it Information in black hole radiation},  Phys. Rev. Lett. {\bf 71} (1993) 3743.
\bibitem{alm2013}
A Almheiri, D Marolf, J Polchinski and J Sully, {\it Black holes: complementarity or firewalls?},  JHEP {\bf 02} (2013), 62.
\bibitem{pol2017}
J Polchinski, {\it The black hole information problem}, in Proceedings , Theoretical advanced study institute in elementaryc particle physics: new frontiers in fields and strings, Boulder, CO, USA (2015) 353.
\bibitem{barrabes2013}
C Barrabes and P. A. Hogan: {\it Advanced general relativity. Gravitational waves, spinning particles and black holes}. Oxford Science Publ., Oxford (2013).
\bibitem{slagter2021}
R J Slagter: {\it A New black hole solution in conformal dilaton gravity on a warped spacetime}, J. Mod. Phys. {\bf 12} (2021), 1758.
\bibitem{san1987}
N Sanchez and B F Whiting:  {\it Quantum field theory and the antipodal identification of black holes},  Nucl. Phys. {\bf B283} (1987) 605.
\bibitem{fol1987}
A Folacci and N Sanchez: {\it Quantum field theory and the antipodal identification of de Sitter space. Elliptic inflation} NASA Astrophysical Data System, paper II.2 (1987).
\bibitem{thooft2015}
G 't Hooft: {\it Diagonalizing the black hole information retrieval process}, arXiv: gr-qc/150901695 (2015).
\bibitem{thooft2016}
G 't Hooft: {\it Black hole unitarity and antipodal entanglement},  Found. Phys. {\bf 46} (2016) 1185.
\bibitem{thooft2018}
G 't Hooft: {\it The firewall transformation for black holes and some of its implications},  arXiv: gr-qc/161208640v3 (2018).
\bibitem{thooft2018a}
G 't Hooft: {\it Discreteness of black hole microstates},  arXiv: gr-qc/180905367v2 (2018).
\bibitem{thooft2018b}
G 't Hooft: {\it Virtual black holes and spacetime structure}, Found. Phys. {\bf 48} (2018), 1149.
\bibitem{thooft2018c}
G 't Hooft: {\it What happens in a black hole when a particle meets its antipode}, arXiv: gr-qc/180405744 (2018).
\bibitem{thooft2019}
G 't Hooft:  {\it The quantum black hole as a theoretical lab}, arXiv: gr-qc/190210469 (2019).
\bibitem{thooft2021}
G 't Hooft: {\it The black hole firewall transformation and realism in quantum mechanics}, arXiv: gr-qc/200611152 (2021) 
\bibitem{groen2020}
N Groenenboom: {\it Quantum gravity on the black hole horizon}, Thesis, Univ. Utrecht (2020).
\bibitem{bet2016}
P. Betzios, N. Gaddam and O. Papadoulaki: {\it The black hole S-matrix from quantum mechanics}, JHEP {\bf 11} (2016), 131.
\bibitem{schrod1957}
E Schr\"odinger: {\it Expanding universe}, Cambridge University. Press, Cambridge U.K. (1957).
\bibitem{thooft2015b}
G 't Hooft: {\it Singularities, horizons, firewalls and local conformal symmetry}, arXiv: gr-qc/151104427 (2015).
\bibitem{codello2013}
A Codello, G D'Odorico, G Pagani and R Percacci, {\it The renormalization group and Weyl-invariance}, Class. Quant. Grav.  {\bf 30} (2013) 115015.
\bibitem{alvarez2014}
E Alvarez, M Herrero-Valea and C P Martin: {\it Conformal and non conformal dilaton gravity, JHEP},  {\bf 10} (2014) 214.
\bibitem{slagter2019c}
R J  Slagter:  {\it Conformal invariant gravity coupled to a gauged scalar field and warped spacetimes}, Phys. Dark Universe,  {\bf 24} (2018) 100282.
\bibitem{slagter2021a}
R J Slagter: {\it Conformal dilaton gravity and warped spacetimes in 5D },
arXiv: gr-qc/201200409 (2021).
\bibitem{felsager1998}
B Felsager: {\it Geometry, particles and fields}, Springer, New York (1998).
\bibitem{slagter2022a}
R J Slagter: {\it The dilaton black hole on a conformal invariant five-dimensional worped spacetime: paradoxes possibly resolved?}, arXiv:gr-qc/220306506(2021).
\bibitem{randall1999a}
L Randall and  R Sundrum:  {\it A large mass hierarchy from a small extra dimension},  Phys. Rev. Lett. {\bf 83}, (1999,) 3370
\bibitem{randall1999b}
L Randall and R Sundrum: {\it An alternative to compactification}, Phys. Rev. Lett. {\bf 83}, (1999) 4690.
\bibitem{ark1998}
N Arkani-Hamed, S Dimopoulos  and G R Dvali: {\it The hierarchy problem and new dimensions at a millimeter}, Phys. Lett. B {bf 429}, (1998) 263.
\bibitem{shirom2000}
T Shiromizu, K Maeda and M Sasaki, {\it The Einstein equations on the 3-brane world}, Phys. Rev. D {\bf 62}, (2000) 024012.
\bibitem{shirom2003}
T Shiromizu, K  Maeda and M Sasaki: {\it Low energy effective theory for two branes system-covariant curvature formulation},  Phys. Rev. D {\bf 7}, (2003) 084022. 
\bibitem{slagterpan2016}
R J Slagter and  S Pan: {\it A new fate of a warped 5D FLRW model with a U(1) scalar gauge field}, Found. of Phys. {\bf 46}, (2016) 1075.
\bibitem{maartens2010}
R Maartens and K Koyama,  {\it Brane-world gravity}, Liv. Rev. Rel. {\bf 13}, (2010), 5. 
\bibitem{banadoz1992}
M Ba\u nados, C. Teitelboim and T. Zanelli: {\it The black hole in three-dimensional spacetime}, Phys. Rev. Lett., {\bf 69} (1992) 1849.
\bibitem{compere2018}
G. Comp\`ere: {\it Advanced lectures on general relativity}, Springer, New York (2018).
\bibitem{slagter2019b}
R J Slagter: {\it On the dynamical 4D BTZ black hole solution in conformally invariant gravity}, J. Mod. Phys. {\bf 11} (2019) 1711.
\bibitem{klein1888}
F Klein: {\it Lectures on the Icosahedron and the solution of equations of the fifth degree}. Tr\"uber and Co. (1888).
\bibitem{Toth2002}
G Toth:  {\it Finite M\"obius groups, minimal immersions of spheres and moduli}. Springer, Heidelberg. (2002).
\bibitem{Shurman1997}
J Shurman: {\it Geometry of the quintic}. John Wiley and Sons Inc., New York. (1997).
\bibitem{king1992}
R B King and F R Canfield: {\it Icosahedral symmetry and the quintic equation},
Computers Math. Applic., {\bf 24},  (1992) 13.
\bibitem{king1996}
R B King: {\it Beyond the quartic equation}. Birkhauser, Boston Inc., Boston (1996).
\bibitem{duke2002}
W Duke and A Toth: {\it The splitting of primes in division fields of elliptic curves}, Exp. Math.   {\bf 11}, (2002) 555.
\bibitem{conn1999}
I Connall: {\it Elliptic curve book}, http://www.math.mcgill.ca/connell/
\bibitem{gauld2006}
D B Gauld:  {\it Differential topology}. Dover Publ., Mineola, New York (2006).
\bibitem{Steenrod1951}
N E Steenrod, {\it Topology of fibre bundles}. Princeton Univ. Press, New York. (1951).
\bibitem{ur2003}
H K Urbantke: {\it The Hopf fibration-seven times in physics}, J. Geom and Phys. {\bf 46}, (2003) 125.
\bibitem{martin2001}
F J Lopez and F Martin: {\it Complete non-orientable minimal surfaces in $\mathds{R}^3$}, Publ. Matematiques {\bf 43} (1999)  341. 
\bibitem{lopez1993}
F J Lopez, {\it A complete minimal Klein bottle in $\mathds{R}^3$}, Duke Math. J. {\bf 71}  (1993) 23. 
\bibitem{pesci2015}
A I Pesci, R E Goldstein, G P Alexander and H Keith Moffatt: {\it Instability of a M\"obius strip minimal surface and a link with systolic geometry}, Phys. Rev. Lett. {\bf 114} (2015) 127801.
\bibitem{hoffman1990}
D Hoffman and W H Meeks, III: {\it The strong half-space theorem for minimal surfaces}, Invent. Math., {\bf 101}, (1990) 373.
\bibitem{penrose1984}
R Penrose and W Rindler:  {\it Spinors and spacetime}. Cambridge Univ. Press. Cambridge. (1984).
\bibitem{pras2001}
V V Prasolov: {\it Algorithms and computations in mathematics, Vol. II}, Springer, New York (2001).
\bibitem{schrod1936}
E Schr\"odinger: {\it Probability relations between separated systems}, Proc. Camb. Soc. {\bf 32} (1936) 446
\bibitem{kirk2006}
K A Kirkpatrick: {\it The Schr\"odinger-HJW theorem}, Found. Phys. Lett. {\bf 19} (2006) 95

\end{thebibliography}
\end{document}